\def\year{2021}\relax
\documentclass[letterpaper]{article} 
\usepackage{aaai21}  
\usepackage{times}  
\usepackage{helvet} 
\usepackage{courier}  
\usepackage[hyphens]{url}  
\usepackage{graphicx} 
\usepackage{natbib}  
\usepackage{caption} 
\usepackage[utf8]{inputenc} 
\usepackage[T1]{fontenc}    
\usepackage{url}            
\usepackage{float}
\usepackage{adjustbox}
\usepackage{booktabs} 
\usepackage{graphicx}
\usepackage{multirow}
\usepackage{ragged2e}
\usepackage{amsmath}
\usepackage[ruled,vlined,linesnumbered]{algorithm2e}
\usepackage{amsfonts}       
\usepackage{nicefrac}       
\usepackage{microtype}      
\usepackage{enumitem}
\usepackage{xcolor,colortbl}
\usepackage[ruled,vlined,linesnumbered]{algorithm2e}
\usepackage{amsfonts}       
\usepackage{nicefrac}       
\usepackage{microtype}      
\usepackage{enumitem}

\definecolor{lightgray}{rgb}{0.83, 0.83, 0.83}
\definecolor{darkgray}{rgb}{0.5, 0.5, 0.5}
\definecolor{mediumgray}{rgb}{0.66, 0.66, 0.66}

\nocopyright
\title{Detecting Emergent Intersectional Biases: \\ Contextualized Word Embeddings Contain a Distribution of Human-like Biases}
\author{
   Wei Guo\textsuperscript{\rm 1} and
    Aylin Caliskan\textsuperscript{\rm 1, 2}\\
}
\affiliations {
    \textsuperscript{\rm 1}George Washington University \\
    \textsuperscript{\rm 1}Department of Computer Science \\
    \textsuperscript{\rm 2}Institute for Data, Democracy \& Politics\\
    weiguo@gwu.edu and aylin@gwu.edu
}

\begin{document}

\maketitle

\makebox[0.45\textwidth][c]{%
    \begin{minipage}{0.42\textwidth}

        \begin{abstract}
With the starting point that implicit human biases are reflected in the statistical regularities of language, it is possible to measure biases in English static word embeddings. State-of-the-art neural language models generate dynamic word embeddings dependent on the context in which the word appears. Current methods measure pre-defined social and intersectional biases that appear in particular contexts defined by sentence templates. Dispensing with templates, we introduce the Contextualized Embedding Association Test (CEAT), that can summarize the magnitude of overall bias in neural language models by incorporating a random-effects model. Experiments on social and intersectional biases show that CEAT finds evidence of all tested biases and provides comprehensive information on the variance of effect magnitudes of the same bias in different contexts. All the models trained on English corpora that we study contain biased representations. GPT-2 contains the smallest magnitude of overall bias followed by GPT, BERT, and then ELMo, negatively correlating with how contextualized the models are.

Furthermore, we develop two methods, Intersectional Bias Detection (IBD) and Emergent Intersectional Bias Detection (EIBD), to automatically identify the intersectional biases and emergent intersectional biases from static word embeddings in addition to measuring them in contextualized word embeddings. We present the first algorithmic bias detection findings on how intersectional group members are strongly associated with unique emergent biases that do not overlap with the biases of their constituent minority identities. IBD achieves an accuracy of 81.6\% and 82.7\%, respectively, when detecting the intersectional biases of African American females and Mexican American females, where the random correct identification rates are  14.3\% and 13.3\%. EIBD reaches an accuracy of 84.7\% and 65.3\%, respectively, when detecting the emergent intersectional biases unique to African American females and Mexican American females, where the random correct identification rates are 9.2\% and 6.1\%. Our results indicate that intersectional biases associated with  members of multiple minority groups, such as African American females and Mexican American females, have the highest magnitude across all neural language models.

\end{abstract}
    \end{minipage}}

\newpage
\section{Introduction}
\label{sec:intro}

State-of-the-art off-the-shelf neural language models such as the multi-million dollar GPT-3, associates men with competency and occupations demonstrating higher levels of education, in downstream natural language processing (NLP) tasks such as sequence prediction \cite{brown2020language}. When GPT-3's user interface for academic access is prompted for language generation with the input ``What is the gender of a doctor,'' the first answer is ``A: Doctor is a masculine noun;'' whereas when prompted with ``What is the gender of a nurse,'' the first answer is ``It’s female.'' Propagation of social group bias in NLP applications such as automated resume screening, that shapes the workforce by making consequential decisions about job candidates, would not only perpetuate existing biases but potentially exacerbate harmful bias in society to affect future generations \cite{de2019bias, raghavanchallenges}. To enhance transparency in NLP, we use the representations of words learned from word co-occurrence statistics to discover social biases.
Our methods uncover unique intersectional biases associated with individuals that are members of multiple minority groups. After identifying these emergent biases, we use numeric representations of words that vary according to neighboring words to analyze how prominent bias is in different contexts. Recent work has shown that human-like biases are embedded in the statistical regularities of language that are learned by word representations, namely word embeddings \cite{caliskan2017semantics, blodgett2020language}. We build a method on this work to automatically identify intersectional biases, such as the ones associated with African American and Mexican American women from static word embeddings (SWE). Then, we measure how human-like biases manifest themselves in contextualized word embeddings (CWE), which are dynamic word representations generated by neural language models that adapt to their context. 

 Artificial intelligence systems are known not only to perpetuate social biases, but they may also amplify existing cultural assumptions and inequalities \cite{campolo2017ai}. While most work on biases in word embeddings focuses on a single social category (e.g., gender, race) \citep{caliskan2017semantics, bolukbasi2016man, garg2018word,zhao2018learning,gonen2019lipstick}, the lack of work on identifying intersectional biases, the bias associated with populations defined by multiple categories \citep{cabreradiscovery}, leads to an incomplete measurement of social biases \citep{hancock2007multiplication,hurtado2008more}. For example, \citet{caliskan2017semantics}'s Word Embedding Association Test (WEAT) quantifies biases documented by the validated psychological methodology of the Implicit Association Test (IAT) \citep{greenwald1998measuring, greenwald2003understanding}. The IAT provides the sets of words to represent social groups and attributes to be used while measuring bias. Consequently, the analysis of bias via WEAT is limited to the types of IATs and their corresponding words contributed by the IAT literature, which happens to include intersectional representation for only African American women. To overcome these constraints of WEATs, we extend WEAT to automatically identify attributes associated with individuals that are members of more than one social group. While this allows us to discover emergent intersectional biases, it is also a promising step towards automatically identifying all biased associations embedded in the regularities of language. To fill the gap in understanding the complex nature of intersectional bias, we develop a method called Intersectional Bias Detection (IBD) to automatically identify intersectional biases without relying on pre-defined attribute sets from the IAT literature.


Biases associated with intersectional group members contain emergent elements that do not overlap with the biases of their constituent minority identities \citep{ghavami2013intersectional,arrington201513}.
 For example, "hair weaves" is stereotypically associated with African American females but not with African Americans or females.
We extend IBD and introduce a method called Emergent Intersectional Bias Detection (EIBD) to identify the emergent intersectional biases of an intersectional group in SWE. Then, we construct new tests to quantify these intersectional and emergent biases in CWE.
To investigate the influence of different contexts, we use a fill-in-the-blank task called masked language modeling. The goal of the task is to generate the most probable substitution for the [MASK] that is surrounded with neighboring context words in a given sentence. BERT, a widely used language model trained on this task, substitutes [MASK] in ``Men/women \textit{excel} in [MASK].'' with ``science'' and ``sports'', reflecting stereotype-congruent associations. However, when we feed in similar contexts ``The man/woman is \textit{known} for his/her [MASK],'' BERT fills ``wit'' in both sentences, which indicates gender bias may not appear in these contexts. Prior methods use templates analogous to masked language modeling to measure bias in CWE \citep{may2019measuring,tan2019assessing,kurita2019quantifying}. The templates are designed to substitute words from WEAT's sets of target words and attributes in a simple manner such as "This is [TARGET]" or "[TARGET] is a [ATTRIBUTE]".
In this work, we propose the Contextualized Embedding Association Test (CEAT), a test eschewing templates and instead generating the distribution of effect magnitudes of biases in different contexts from a control corpus. To comprehensively measure the social and intersectional biases in this distribution, a random-effects model designed to combine effect sizes of similar bias interventions summarizes the overall effect size of bias in the neural language model \citep{dersimonian2007random}. As a result, instead of focusing on biases template-based contexts, CEAT measures the distribution of biased associations in a language model.

\noindent \textbf{Contributions.} In summary, this paper presents three novel contributions along with three complementary methods (CEAT, IBD, and EIBD) to automatically identify intersectional biases as well as emergent intersectional biases in SWE, then use these findings to measure all available types of social biases in CWE. We find that ELMo is the most biased, followed by BERT, then GPT, with GPT-2 being the least biased. The overall level of bias correlated with how contextualized the CWE generated by the models are. Our results indicate that the strongest biased associations are embedded in the representations of intersectional group members such as African American women. Data, source code, and detailed results are available.

\noindent  \textbf{Intersectional Bias Detection (IBD).} We develop a novel method for SWE to detect words that represent biases associated with intersectional group members. To our knowledge, IBD is the first algorithmic method to automatically identify individual words that are strongly associated with intersectionality. IBD reaches an accuracy of 81.6\% and 82.7\%, respectively, when evaluated on intersectional biases associated with African American females and Mexican American females that are provided in \citet{ghavami2013intersectional}'s validation dataset. In these machine learning settings, the random chances of correct identification are 14.3\% and 13.3\%. Currently, the validation datasets represent gender as a binary label. Consequently, our method uses binary categorization when evaluating for gender related biases. However, we stress that our method generalizes to multiple categories from binary. In future work, we aim to design non-categorical methods that don't represent individuals as members of discrete categories compared to potentially using continuous representations. Accordingly, we also plan to compile validation datasets that won't constrain our evaluation to categorical assumptions about humans. 
 
 \noindent \textbf{Emergent Intersectional Bias Detection (EIBD).} We contribute a novel method to identify emergent intersectional biases that do not overlap with biases of constituent social groups in SWE. To our knowledge, EIBD is the first algorithmic method to detect the emergent intersectional biases in word embeddings automatically. EIBD reaches an accuracy of 84.7\% and 65.3\%, respectively, when validating on the emergent intersectional biases of African American females and Mexican American females that are provided provided in \citet{ghavami2013intersectional}'s validation dataset. In these machine learning settings, the random chances of correct identification are 9.2\% and 6.1\%. 
 
 \noindent \textbf{Contextualized Embedding Association Test (CEAT).} WEAT measures human-like biases in SWE. We extend WEAT to the dynamic setting of neural language models to quantify the distribution of effect magnitudes of social and intersectional biases in \textit{contextualized} word embeddings and summarize the combined magnitude of bias by pooling effect sizes with the validated random-effects methodology \cite{hedges1983random, borenstein2007meta}. We show that the magnitude of bias greatly varies according to the context in which the stimuli of WEAT appear. Overall, the pooled mean effect size is statistically significant in all CEAT tests including intersectional bias measurements and all models contain biased representations.
\section{Related Work}
\label{sec:related}
SWE are trained on word co-occurrence statistics of corpora to generate numeric representations of words so that machines can process language \citep{mikolov2013distributed,pennington2014glove}. Previous work on bias in SWE has shown that human-like biases that have been documented by the IAT are embedded in the statistical regularities of language \citep{caliskan2017semantics}. The IAT \citep{greenwald1998measuring} is a widely used measure of implicit bias in human subjects that quantifies the differential reaction time to pairing two concepts. Analogous to the IAT,  \citet{caliskan2017semantics} developed the WEAT to measure the biases in SWE by quantifying the relative associations of two sets of target words (e.g., African American and European American) that represent social groups with two sets of polar attributes (e.g., pleasant and unpleasant). WEAT computes an effect size (Cohen's $d$) that is a standardized bias score and its $p$-value based on a one-sided permutation test. WEAT measures biases pre-defined by the IAT such as racism, sexism, ableism, and attitude towards the elderly, as well as widely shared non-discriminatory non-social group associations. \citet{swinger2019biases} presented an adaptation of the WEAT to identify biases associated with clusters of names.

Regarding the biases of intersectional groups categorized by multiple social categories, there is prior work in the social sciences focusing on the experiences of African American females \citep{crenshaw1989demarginalizing,hare1988meaning, kahn1989psychology,thomas1995psychology}. Buolamwini et al. demonstrated intersectional accuracy disparities in commercial gender classification in computer vision \citep{buolamwini2018gender}. \citet{may2019measuring} and \citet{tan2019assessing} used the attributes presented in \citet{caliskan2017semantics} to measure emergent intersectional biases of African American females in CWE. We develop the first algorithmic method to automatically identify intersectional bias and emergent bias attributes in SWE, which can be measured in both SWE and CWE. Furthermore, we construct new embedding association tests for the intersectional groups. As a result, our work is the first to discuss biases regarding Mexican American females in word embeddings. \citet{ghavami2013intersectional} used a free-response procedure in human subjects to collect words that represent intersectional biases. They show that emergent intersectional biases exist in several gender-by-race groups in the U.S. We use the validation dataset constructed by \citet{ghavami2013intersectional} to evaluate our methods.

Recently, neural language models, which use neural networks to assign probability values to sequences of words, have achieved state-of-the-art results in NLP tasks with their dynamic word representations, CWE \citep{edunov2018understanding,bohnet2018morphosyntactic,yang2019xlnet}. Neural language models typically consist of an encoder that generates CWE for each word based on its accompanying context in the input sequence. Specifically, the collection of values on a particular layer's hidden units forms the CWE \citep{tenney2019you}, which has the same vector shape as a SWE. However, unlike SWE that represent each word, including polysemous words, with a fixed vector, CWE of the same word vary according to its context window that is encoded into its representation by the neural language model.  \citet{ethayarajh2019understanding} demonstrate how these limitations of SWE impact measuring gender biases. With the wide adaption of neural language models \citep{edunov2018understanding,bohnet2018morphosyntactic,yang2019xlnet}, human-like biases were observed in CWE \citep{kurita2019quantifying,zhao2019gender,may2019measuring,tan2019assessing}.
 To measure human-like biases in CWE, \citet{may2019measuring} applied the WEAT to contextualized representations in template sentences. \citet{tan2019assessing} adopted the method of \citet{may2019measuring} by applying \citet{caliskan2017semantics}'s WEAT to the CWE of the stimuli tokens in templates such as ``This is a [TARGET]''. \citet{kurita2019quantifying} measured biases in BERT based on the prediction probability of the attribute in a template that contains the target and masks the attribute, e.g., [TARGET] is [MASK].
 \citet{hutchinson2020social} reveal biases associated with disabilities in CWE and demonstrate undesirable biases towards mentions of disability in applications such as toxicity prediction and sentiment analysis. 

%


\citet{nadeem2020stereoset} present a large-scale natural language dataset in English to measure stereotypical biases in the domains of gender, profession, race, and religion. Their strategy cannot be directly compared to ours since it is not aligned with our intersectional bias detection method, which is complementary to CEAT.
 The majority of prior work measures bias in a limited selection of contexts to report the unweighted mean value of bias magnitudes, which does not reflect the scope of contextualization of biases embedded in a neural language model. 

\section{Data}
\label{sec:data}
Identifying and measuring intersectional and social biases in word embeddings as well as neural language models requires four types of data sources that are detailed in this section. (1) SWE carry the signals for individual words that have statistically significant biased associations with social groups and intersectionality. Application of our methods IBD and EIBD to SWE automatically retrieves biased associations. (2) CWE extracted from sentence encodings of neural language models provide precise word representations that depend on the context of word occurrence. We apply CEAT to summarize magnitude of bias in neural language models. (3) A corpus provides the samples of sentences used in CEAT when measuring the overall bias and analyzing the variance of contexts in CWE of neural language models.  (4) Stimuli designed by experts in social psychology represent validated concepts in natural language including social group and intersectional targets in addition to their corresponding attributes.

\subsection{Static Word Embeddings (SWE)}
We use GloVe \cite{pennington2014glove} SWE trained on the word co-occurrence statistics of the Common Crawl corpus to automatically detect words that are highly associated with intersectional group members. The Common Crawl corpus consists of 840 billion tokens and more than 2 million unique vocabulary words collected from a crawl of the world wide web. Consequently, GloVe embeddings capture the language representation of the entire Internet population that contributed to its training corpus. GloVe embeddings learn fine-grained semantic and syntactic regularities  \cite{pennington2014glove}. \citet{caliskan2017semantics} have shown that social biases are embedded in the linguistic regularities learned by GloVe. 

\subsection{Contextualized Word Embeddings (CWE)}

We generate the CWE by widely used neural language model implementations of ELMo from \url{https://allennlp.org/elmo}, BERT, GPT and GPT-2 from \url{https://huggingface.co/transformers/v2.5.0/model_doc/} \cite{peters2018deep,devlin2018BERT,radford2018improving,radford2019language}.  Specifically, CWE  is formed by the collection of values on a particular layer’s hidden units in the neural language model. BERT, GPT and GPT-2 use subword tokenization.
Since GPT and GPT-2 are unidirectional language models, CWE of the last subtokens contain the information of the entire word \cite{radford2019language}. We use the CWE of the last subtoken in the word as its representation in GPT and GPT-2. For consistency, we use the CWE of the last subtoken in the word as its representation in BERT.
BERT and GPT-2 provide several versions. We use BERT-small-cased and GPT-2-117m trained on cased English text. The sizes of the training corpora detailed below have been verified from \citet{assenmacher2020comparability}. We obtained academic access to GPT-3's API which does not provide training data or the CWE. Accordingly, we are not able to systematically study GPT-3.

\textbf{ELMo} is a 2-layer bidirectional long short term memory (Bi-LSTM) \cite{hochreiter1997long} language model trained on the Billion Word Benchmark dataset \cite{chelba2013one} that takes up $\sim$9GB memory. ELMo has 93.6 million parameters. It is different from the three other models since CWE in ELMo integrate the hidden states in all layers instead of using the hidden states of the top layer. 
We follow standard usage and compute the summation of hidden units over all aggregated layers of the same token as its CWE \cite{peters2018deep}. CWE of ELMo have 1,024 dimensions.

\textbf{BERT} \cite{devlin2018BERT} is a bidirectional transformer encoder \cite{vaswani2017attention} trained on a masked language model and next sentence prediction. BERT is trained on BookCorpus \cite{zhu2015aligning} and English Wikipedia dumps  that take up $\sim$16GB memory \cite{bender2021dangers}. We use BERT-small-case with 12 layers that has 110 million parameters. We extract the values of hidden units on the top layer corresponding to the  token as its CWE of 768 dimensions. 

\textbf{GPT} \cite{radford2018improving} is a 12-layer transformer decoder trained on a unidirectional language model on BookCorpus  that takes up $\sim$13GB memory \cite{zhu2015aligning}. We use the values of hidden units on the top layer corresponding to the  token as its CWE. This implementation of GPT has 110 million parameters. The CWE have 768 dimensions. 

\textbf{GPT-2} \cite{radford2019language} is a  transformer decoder trained on a unidirectional language model and is a scaled-up version of GPT. GPT-2 is trained on WebText that takes up $\sim$40GB memory \cite{radford2019language}.
We use GPT-2-small which has 12 layers and 117 million parameters. 
We use the values of hidden units on the top layer corresponding to the   token as its CWE. CWE of GPT-2 have 768 dimensions.

We provide the source code, detailed information, and documentation in our open source repository at \url{https://github.com/weiguowilliam/CEAT}.

\subsection{Corpus}
 We need a comprehensive representation of all contexts a word can appear in naturally occurring sentences in order to investigate how bias associated with individual words varies across contexts. Identifying the potential contexts in which a word can be observed is not a trivial task. Consequently, we simulate the distribution of contexts a word appears in, by randomly sampling sentences that the word occurs in a large corpus.


\citet{voigt2018rtgender} have shown that social biases are projected into Reddit comments. 
Consequently, we use a Reddit corpus to generate the distribution of contexts that words of interest appear in. The corpus consists of 500 million comments made in the period between 1/1/2014 and 12/31/2014.
We take all the stimuli used in \citet{caliskan2017semantics}'s WEAT that measures effect size of bias for social groups and related attributes. For each WEAT type, we retrieve the sentences from the Reddit corpus that contain one of these stimuli. In this way, we collect a great variety of CWE from the Reddit corpus to measure bias comprehensively in a neural language model while simulating the natural distribution of contexts in language. We discuss the justification of sampling 10,000 sentences from the Reddit corpus in the upcoming sections.

\subsection{Stimuli}
\label{subsec:stimuli}
\citet{caliskan2017semantics}'s WEAT is inspired by the IAT literature \cite{greenwald1995implicit, greenwald1998measuring, greenwald2003understanding} that measures implicit associations of concepts by representing them with stimuli. Experts in social psychology and cognitive science select stimuli which are words typically representative of various concepts. These linguistic or sometimes picture-based stimuli are proxies to overall representations of concepts in cognition. Similarly, in the word embedding space, WEAT uses these unambiguous stimuli as semantic representations to study biased associations related to these concepts. Since the stimuli are chosen by experts to most accurately represent concepts, they are not polysemous or ambiguous words. Each WEAT, designed to measure a certain type of association or social group bias, has at least 32 stimuli. There are 8 stimuli for each one of the four concepts. Two of these concepts represent target groups and two of them represent polar attributes. WEAT measures the magnitude of bias by quantifying the standardized differential association or targets with attributes. The larger the set of appropriate stimuli to represent a concept, the more statistically significant and accurate the representation becomes \cite{caliskan2017semantics}.

\noindent \textbf{Validation data for intersectional bias.} To investigate intersectional bias with respect to race and gender, we represent members of social groups with target words provided by WEAT and Parada et al. \citep{caliskan2017semantics,parada2016ethnolinguistic}. WEAT and Parada et al. represent racial categories with frequent given names that signal group membership. WEAT contains a balanced combination of common female and male names of African Americans and European Americans whereas Parada et al. presents the Mexican American names for women and men combined. 
The intersectional bias detection methods identify attributes that are associated with these target group representations. Human subjects provide the validation set of intersectional attributes with ground truth information in prior work \citep{ghavami2013intersectional}. The evaluation of intersectional bias detection methods uses this validation set. One limitation of these validation sets is the way they represent gender as a binary category. We will address this constraint in future work by constructing our own validation sets that won't have to represent people by discrete categorical labels of race and gender.

\section{Approach}
\label{sec:approach}
Our approach includes four components. (1) \cite{caliskan2017semantics}'s WEAT for SWE is the foundation of our approach to summarizing overall bias in CWE generated by neural language models. (2) Random-effects models from the meta analysis literature summarizes the combined effect size for a neural language model's CWE via combining 10,000 WEAT samples by weighting each result with the within-WEAT and between-WEAT variances~\cite{hedges1983random}. (3) Our novel method IBD automatically detects words associated with intersectional biases. (4) Our novel method EIBD automatically detects words that are uniquely associated with members of multiple minority or disadvantaged groups, but do not overlap with the biases of their constituent minority identities. 

Supplementary materials includes the details of all the bias types studied in this paper, namely, WEAT biases introduced by \citet{caliskan2017semantics} as well as intersectional biases and their validation set introduced by \citet{ghavami2013intersectional} and \citet{parada2016ethnolinguistic}.

\subsection{Word Embedding Association Test (WEAT)}
WEAT, designed by \citet{caliskan2017semantics}, measures the effect size of bias in SWE, by quantifying the relative associations of two sets of target words (e.g., career, professional; and family, home) with two sets of  polar attributes (e.g., woman, female; and man, male). Two of these WEATs measure baseline associations that are widely accepted such as the attitude towards flowers vs. insects or the attitude towards musical instruments vs. weapons. Human subjects and word embeddings tend to associate flowers and musical instruments with pleasantness that corresponds to positive valence. However, human subjects associate insects and weapons with unpleasantness that corresponds to negative valence. \citet{greenwald1998measuring} refers to these as universally accepted stereotypes since they are widely shared across human subjects and are not potentially harmful to society. However, the rest of the tests measure the magnitude of social-group associations, such as gender and race stereotypes and attitude towards the elderly or people with disabilities. Biased social-group associations in word embeddings can potentially be prejudiced and harmful to society. Especially, if downstream applications of NLP that use static or dynamic word embeddings to make consequential decisions about individuals, such as resume screening for job candidate selection, perpetuate existing biases to eventually exacerbate historical injustices \cite{de2019bias, raghavanchallenges}.  The formal definition of \citet{caliskan2017semantics}'s WEAT, the test statistic, and the statistical significance of biased associations are detailed in the appendices.

\subsection{Intersectional Bias Detection (IBD) }
IBD identifies words associated with intersectional group members, defined by two social categories simultaneously. Our method automatically detects the attributes that have high associations with the  intersectional group from a set of SWE. Analogous to the Word Embedding Factual Association Test (WEFAT)  \citep{caliskan2017semantics},  we measure the standardized differential association of a single stimulus $w \in W$ with two social groups $A$ and $B$ using the following statistic.
\vspace{-2mm}
\[ s(w, A, B) = \frac{\textrm{mean}_{a \in A} \textrm{cos}(\vec{w}, \vec{a}) - \textrm{mean}_{b \in B} \textrm{cos}(\vec{w}, \vec{b})}{\textrm{std-dev}_{x \in A \cup B}\textrm{cos}(\vec{w}, \vec{x})}\]
\vspace{-2mm}

We refer to the above statistic as the \textbf{association score}, which is used by WEFAT to verify that gender statistics are embedded in linguistic regularities. Targets $A$ and $B$ are words that represent males (e.g., he, him) and females (e.g., she, her) and $W$ is a set of occupations. For example, \textit{nurse} has an association score $s(nurse, A, B)$ that measures effect size of gender associations. WEFAT has been shown to have high predictive validity ($\rho=0.90$) in quantifying facts about the world \citep{caliskan2017semantics}. 

We extend WEFAT's {\em gender} association measurement to quantify the relative association to other social categories (e.g., race), by following an approach similar to lexicon induction that quantifies certain associations without annotating large-scale ground truth training data \cite{hatzivassiloglou1997predicting, riloff2003learning, turney2003measuring}. Let $P_i = (A_i,B_i$) (e.g.,  African American and European American) be a pair of social groups, and $W$ be a set of attribute words. 
We calculate the association score $s(w,A_i,B_i)$ for $w \in W$. If $s(w,A_i,B_i)$ is greater than the positive effect size threshold $t$,  $w$ is detected to be  associated with group $A_i$.
Let $W_i = \{w|s(w,A_i,B_i)>t, w \in W\}$  be the associated word list for each pair $P_i$. 

We detect the biased attributes associated with an intersectional group $C_{mn}$ defined by two  social categories $C_{1n}, C_{m1}$ with $M$ and $N$ subcategories ($C_{11},  \dots, C_{mn}$) (e.g., African American females by race ($C_{1n}$) and gender ($C_{m1}$)). We assume, there are three racial categories $M =3$, and two gender categories $N=2$ in our experiments because of the limited structure of representation for individuals in the validation dataset as well as the stimuli. We plan to extend these methods to non-binary individuals and non-categorical representations. However, precisely validating such an approach would require us to construct the corresponding validation sets, which currently don't exist. \textbf{Generalizing the method to represent humans with continuous values as opposed to categorical group labels is left to future work.}  There are in total $ M \times N $ combinations of intersectional groups $C_{mn}$. We use all  groups $C_{mn}$  to build WEFAT pairs
$P_{ij} = (C_{11}, C_{ij}), i = 1,...,M, j = 1,...,N$. Then, we detect lists of words associated with each pair $W_{ij}, i = 1,...,M, j = 1,...,N$ based on threshold $t$ determined by an ROC curve. We detect the attributes highly associated with the intersectional group, for example C$_{11}$, from all $( M\times N)$ WEFAT pairs.
We define the words associated with intersectional biases of group C$_{11}$ as $W_{IB}$ and these words are identified by 
\vspace{-3mm}

\[W_{IB} = \bigcup_{\substack{1\leq i\leq M\\1\leq j\leq N}}W_{IB_{ij}},\;
\] 
where 
\vspace{-5mm}
 \[ \hspace{12mm} W_{IB_{ij}} = \{w|s(w,C_{11},C{_{ij}})>t_{mn}, w \in W_{IB_{mn}}\} \] 

\noindent where 
\vspace{-3mm}
\[ W_{IB_{mn}} = \{(\bigcup_{\substack{1\leq i\leq M\\1\leq j\leq N}}W_{ij})\cup W_{random}\} \] 

\noindent W$_{11}$ contains validated words associated with C$_{11}$. Each W$_{ij}$ contains validated words associated with one intersectional group \cite{ghavami2013intersectional}. W$_{random}$ contains random words, which are stimuli taken from WEAT that are not associated with any C$_{ij}$, thus represent true negatives. 


To identify the thresholds, we treat IBD as a one-vs-all verification classifier in machine learning to determine whether attributes belong to group $C_{11}$. 
We select the  threshold  with the highest value of $true\: positive\: rate - false\: positive\: rate$ ($TPR - FPR$). When multiple thresholds have the same values, we select the one with the highest $TP$ to detect more attributes associated with $C_{11}$. Detection accuracy is calculated as true positives plus true negatives over true positives plus true negatives plus false positives plus false negatives $(\frac{TP+TN}{TP+TN+FP+FN})$. The attributes which are associated with $C_{11}$ and are detected as $C_{11}$ are $TP$. The attributes which are not associated with $C_{11}$ and are not detected as $C_{11}$ are $TN$.  The attributes which are associated with $C_{11}$ but are not detected as $C_{11}$ are $FN$. The attributes which are not associated with $C_{11}$ but are detected as $C_{11}$ are $FP$.

\subsection{Emergent Intersectional Bias Detection (EIBD)}
EIBD identifies words that are uniquely associated with intersectional group members. These emergent biases are only associated with the intersectional group (e.g., African American females $C_{11}$) but not associated with its constituent category such as African Americans $S_{1n}$ or females $S_{m1}$. EIBD is a modified and extended version of IBD. The formal definition is in the appendices.


Conceptually, to detect words uniquely associated with African American females in a set of attributes $W$, we assume there are  two classes (females, males) of gender and two classes (African Americans, European Americans) of race.
We measure the relative association of all words in $W$ first
with African American females and African American males, second with African American females and European American females, third with African American females and European American males. (Fourth is the comparison of the same groups, which leads to $d=0$ effect size, which is always below the detection threshold.) The union of attributes with an association score greater than the selected threshold represents intersectional biases associated with African American females. 
Then, we calculate the association scores of these IBD attributes first with females and males, second with African Americans and European Americans. We remove the attributes with scores greater than the selected threshold from these IBD attributes, that are highly associated with single social categories. The union of the remaining attributes are the emergent intersectional biases.

\subsection{Contextualized Embedding Association Test (CEAT)}
CEAT quantifies social biases in CWE by extending the WEAT methodology that measures human-like biases in SWE \citep{caliskan2017semantics}. 
WEAT's bias metric is effect size (Cohen's $d$). In CWE, since  embeddings of the same word vary based on context, applying WEAT to a biased set of CWE will not measure bias comprehensively. To deal with a range of dynamic embeddings representing individual words, CEAT measures the distribution of  effect sizes that are embedded in a neural language model. 

In WEAT's formal definition  \citep{caliskan2017semantics}, $X$ and $Y$ are two sets of target words of equal size; $A$ and $B$ are two sets of evaluative polar attribute words of equal size. Each word in these sets of words is referred to as a stimulus. Let $cos(\vec{a},\vec{b})$ stand for the cosine similarity between vectors $\vec{a}$ and $\vec{b}$. 
WEAT measures the magnitude of bias by computing the effect size ($ES$) which is the standardized differential association of the targets and attributes. The $p$-value ($P_w$) of WEAT measures the probability of observing the effect size in the null hypothesis, in case biased associations did not exist. According to Cohen's effect size metric, $d > \mid 0.5 \mid$ and $d > \mid 0.8\mid$ are medium and large effect sizes, respectively  \citep{rice2005comparing}.


In a neural language model, each stimulus $s$ from WEAT contained in $n_s$ input sentences has at most $n_s$ different CWE $\vec{s_1},..., \vec{s_{n_s}}$ depending on the context in which it appears.
If we calculate effect size $ES(X,Y,A,B)$ with all different $\vec{s}$ for a stimulus $s \in X$ and keep the CWE for other stimuli unchanged, there will be at most $n_s$ different values of effect size. For example, if we assume each stimulus $s$ occurs in 2 contexts and each set in $X, Y, A, B$  has 5 stimuli, the total number of combinations for all the CWE of stimuli will be $2^{5\times4} = 1,048,576$. The numerous possible values of $ES(X,Y,A,B)$ construct a  \textit{distribution} of effect sizes, therefore we extend WEAT to CEAT.


For each CEAT, all the sentences, where a CEAT stimulus occurs, are retrieved from the Reddit corpus. Then, we generate the corresponding CWE from these sentences with randomly varying contexts. In this way, we generate $n_s$ CWE from $n_s$ extracted sentences for each stimulus $s$, where $n_s$ can vary according to the contextual variance of each stimulus.
We sample random combinations of CWE for each stimulus $N$ times. In the $i^{th}$ sample out of $N$, for each stimulus that appears in at least $N$ sentences, 
we randomly sample one of its CWE vectors without replacement. If a stimulus occurs in less than $N$ sentences, especially when $N$ is very large, we randomly sample from its CWE vectors with replacement so that they can be reused while preserving their distribution. We provide the analysis and extended results in the appendices for both $N=1,000$ and $N=10,000$, which result in similar bias magnitudes. Based on the sampled CWEs, we calculate each sample's effect size $ES_i(X,Y,A,B)$, sample variance $V_i(X,Y,A,B)$  and $p$-value $P_{w_i}(X,Y,A,B)$  in WEAT. Then, we generate $N$ of these samples to approximate the distribution of effect sizes via CEAT.





The distribution of bias effects in CEAT represents random effects computed by WEAT where we do not expect to observe the same effect size due to variance in context \cite{hedges1983random}. As a result, in order to provide comprehensive summary statistics, we applied a random-effects model from the validated meta-analysis literature to compute the weighted mean of the effect sizes and statistical significance \citep{rosenthal2002meta, borenstein2007meta}. The summary of the effect magnitude of a particular bias in a neural language model, namely combined effect size (CES), is the weighted mean of a distribution of random effects,
\vspace{-1mm}
\[CES(X,Y,A,B) = \frac{\sum_{i=1}^{N}v_i ES_i}{\sum_{i=1}^{N}v_i}\]
\vspace{-2mm}

\noindent where $v_i$ is the inverse of the sum of in-sample variance $V_i$ and between-sample variance in the distribution of random effects $\sigma_{between}^2$.  Methodological details are in the appendices.

\subsection{Random-Effects Model}
\label{subsec:random}

Meta-analysis is the statistical procedure for combining data from multiple studies \cite{hedges1998fixed}. Meta-analysis  describes the results of each separate study by a numerical index (e.g., effect size) and then summarizes the results into combined statistics. In bias measurements, we are dealing with effect size. Based on different assumptions whether the effect size is fixed or not, there are two kinds of methods: \textit{fixed-effects} model and \textit{random-effects} model. 
Fixed-effects model expects results with fixed-effect sizes from different intervention studies. On the other hand, random-effects model treats the effect size as they are samples from a random distribution of all possible effect sizes \cite{dersimonian1986meta,hedges2014statistical}. The expected results of different intervention studies in the random-effects model don't have to match other studies' results. 
In our case, since the effect sizes calculated with the CWE in different contexts are expected to vary, we cannot assume a fixed-effects model. Instead, we use a random-effects model that is appropriate for the type of data we are studying. 

We apply a random-effects model from the validated meta-analysis literature using the methods of \citet{hedges1998fixed}. Specifically, we describe the procedures for estimating the comprehensive summary statistic, \textbf{combined effect size (CES)}, which is the weighted mean of a distribution of random-effect sizes. Each effect size is weighted by the variance in calculating that particular effect size in addition to the overall variance among all the random-effect sizes. 

We combine effect size estimates from $N$ independent WEATs. The details of CES are in the appendices.

\begin{figure*}[ht!]
  \centering
  {%
\includegraphics[width=0.8\textwidth]{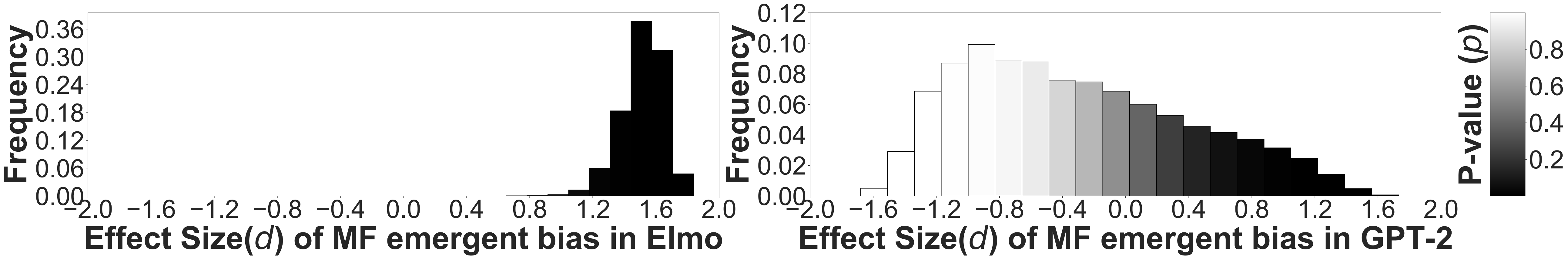}
  }
 
  \caption{\small{Distributions of effect sizes with ELMo (CES $d=1.51$) and GPT-2 (CES $d=-0.32$) for emergent intersectional bias CEAT test I4. Test I4, after identifying the emergent and intersectional biases associated with Mexican American females and European American males (MF/EM) via IBD and EIBD in word embeddings, CEAT measures the overall distribution of biased associations for the retrieved stimuli in the neural language models.
This example is chosen to demonstrate how different models exhibit varying degrees of bias when using the same set of stimuli to measure bias. The height of each bar shows the frequency of observed effect sizes among 10,000 effect size samples of a particular bias type that fall in each bin. The color coded bars stand for the average $p$-value of all effect sizes corresponding to that bin.}}
\label{fig:weat}
\end{figure*}




 \begin{figure*}[ht!]
  \centering
  {%
\begin{tabular}{cccc}
  \includegraphics[width=.2\textwidth]{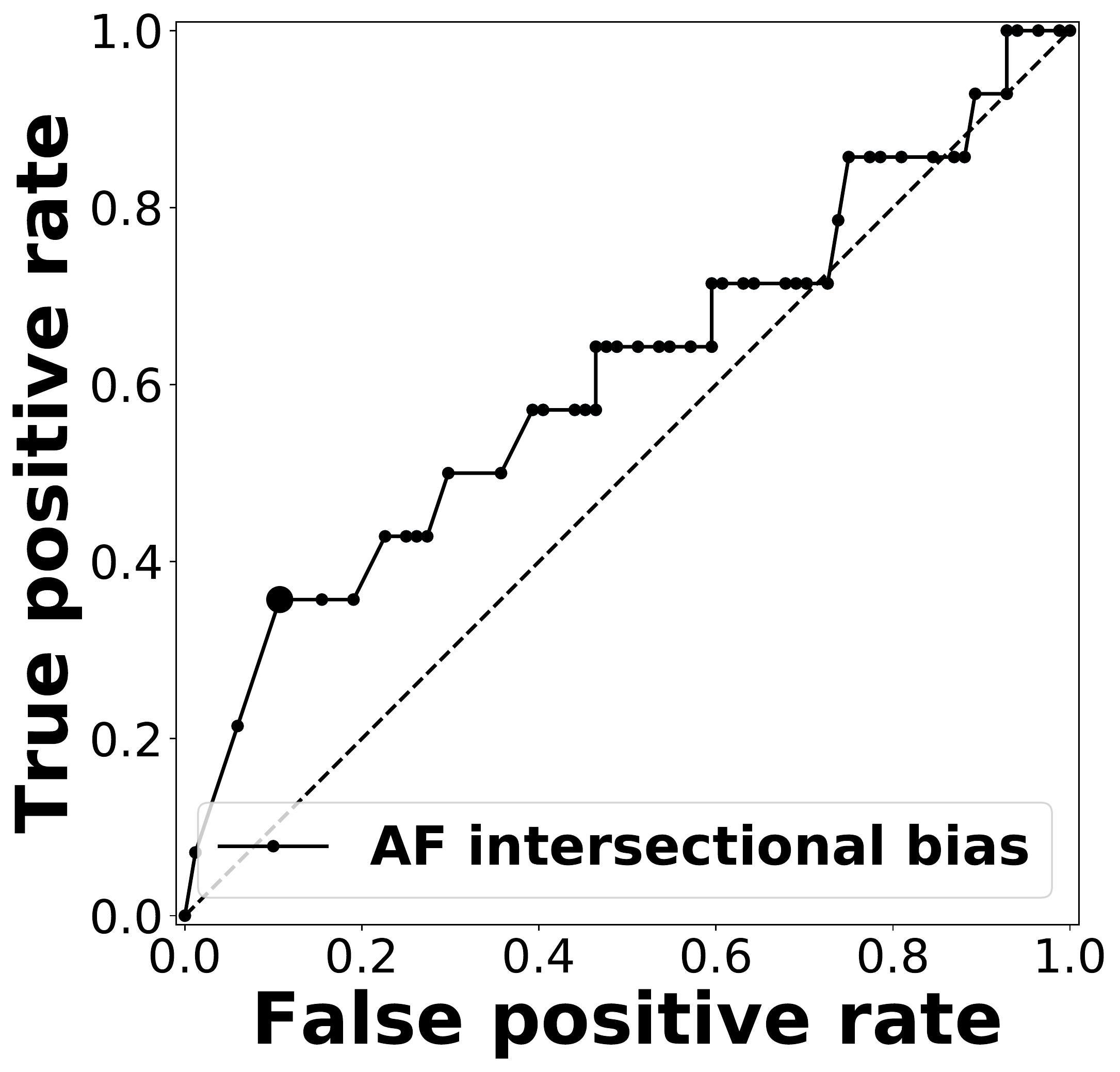} &
    \includegraphics[width=.2\textwidth]{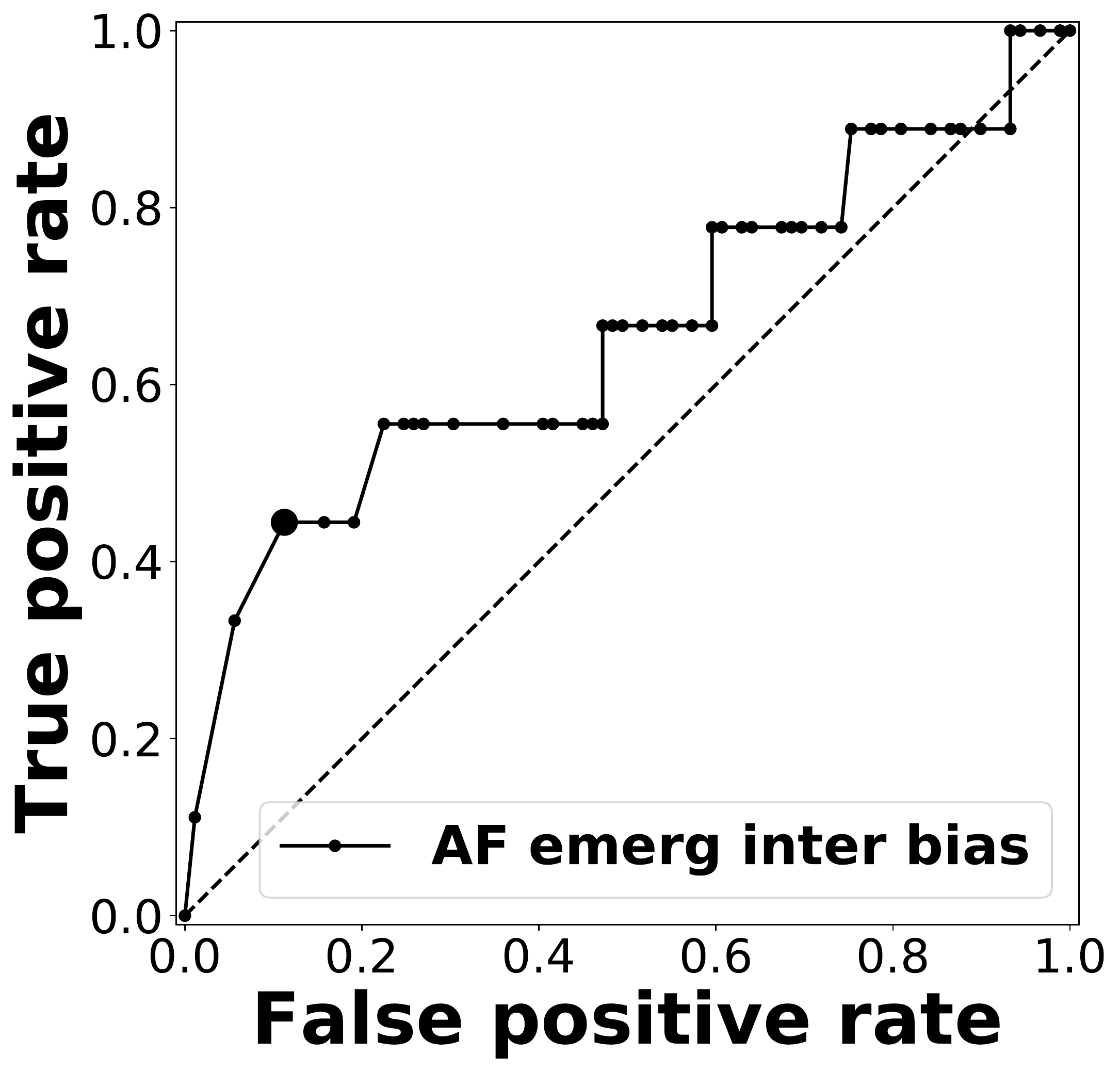} &
      \includegraphics[width=.2\textwidth]{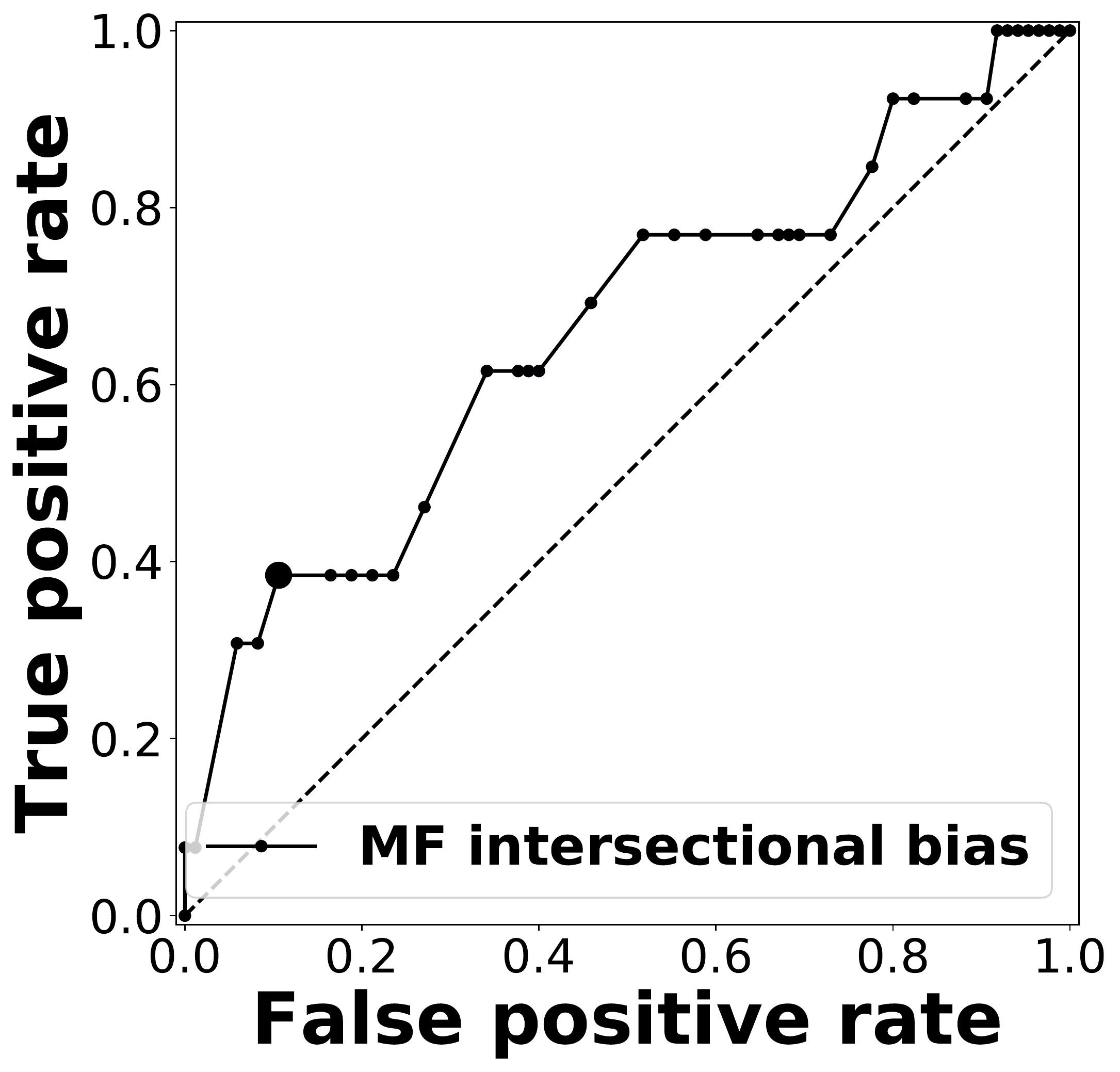} &
\includegraphics[width=.2\textwidth]{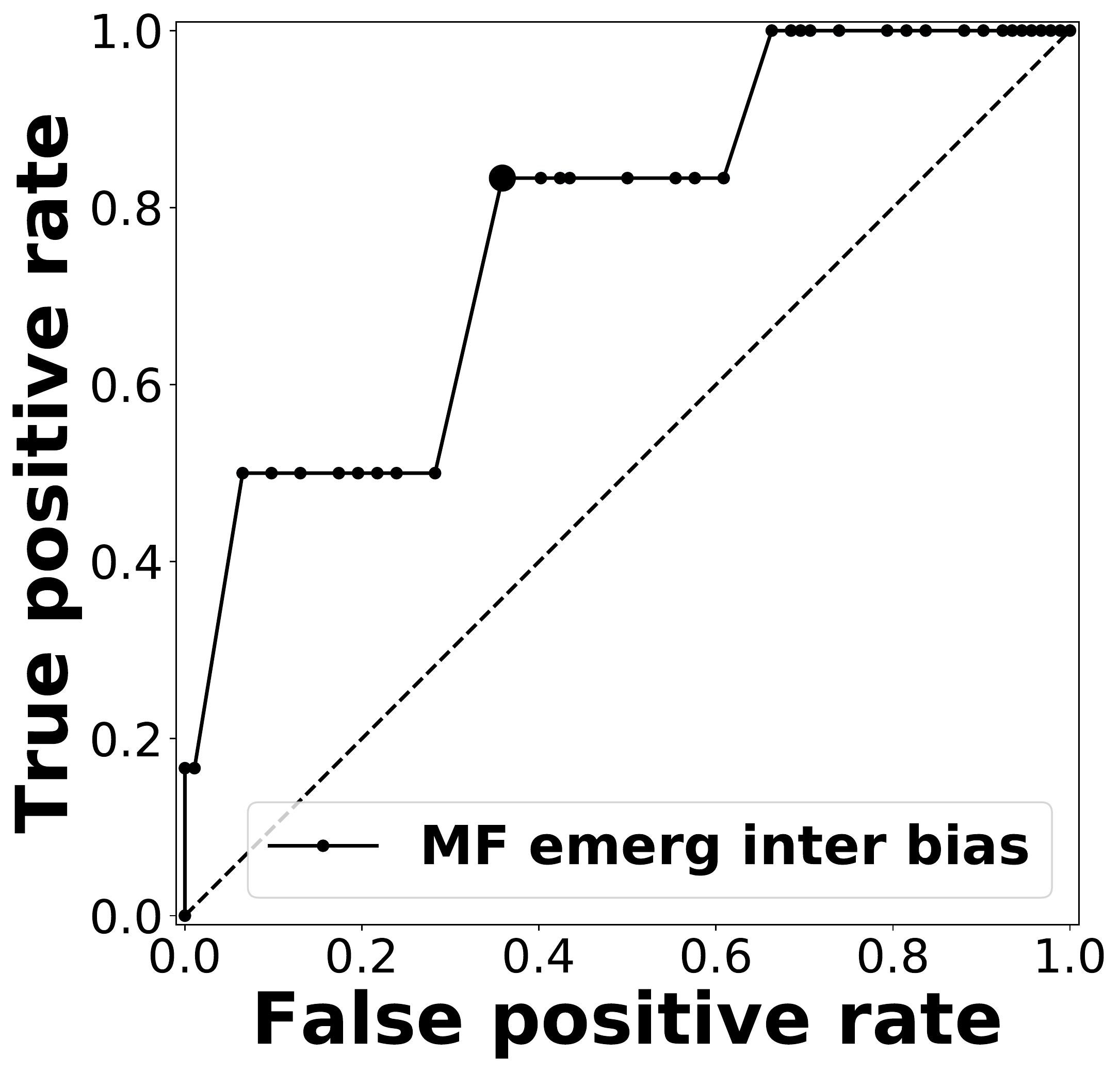}\\
  \end{tabular}}
  \vspace{-2mm} \caption{\textbf{ROC curves of IBD and EIBD for African American females (AF) and Mexican American females (MF).} The value that maximizes the $true\: positive\: rate\: -\: false\: positive\: rate$ is selected as the optimal threshold marked with a dot.
  `emerg inter bias' stands for emergent intersectional bias. 
\vspace{-4mm}  }
  \label{fig:roc}
\end{figure*}

\addtolength{\textfloatsep}{-0.05in}

\section{Results and Evaluation}
\label{sec:experiments}

We measure ten types of social biases via WEAT (C1-C10) and construct our own intersectional bias tests in ELMo, BERT, GPT, and GPT-2. Accordingly, we present four novel intersectional bias tests via IBD and EIBD for studying African American, European American, and Mexican American men and women. 

We use the stimuli introduced in Section~\ref{subsec:stimuli} to represent the target groups. For intersectional and emergent bias tests, we use the attributes associated with the intersectional minority or disadvantaged group members vs the majority European American males as the two polar attribute sets. We sample $N=10,000$ combinations of CWE for each CEAT since according to various evaluation trials, the resulting CES and $p$-value remain consistent under this parameter. 

\subsection{Evaluation of IBD and EIBD}
\label{sec:evaluation}

 We use IBD and EIBD to automatically detect and retrieve the intersectional and emergent biases associated with intersectional group members (e.g., African American females, Mexican American females) in GloVe SWE. 
To evaluate our methods IBD and EIBD, we use validated stimuli provided in prior work that represents each social group with frequent given names, as explained in Section~\ref{sec:data}. 
IBD and EIBD experiments use the same test set consisting of 98 attributes associated with 2 groups defined by gender (females, males), 3 groups defined by race (African American, European American, Mexican American), 6 intersectional groups in total defined by race and gender, in addition to random words taken from WEAT not associated with any group \cite{ghavami2013intersectional}. These random words represent the true negatives for evaluating the identification task.

We draw the ROC curves of four bias detection tasks in Figure~\ref{fig:roc}, then select the highest value of
$TPR - FPR$ as thresholds for each intersectional group. 
IBD achieves an accuracy of 81.6\% and 82.7\%, respectively, when detecting the intersectional biases of African American females and Mexican American females, where the random correct identification rates are 14.3\% and 13.3\%. EIBD reaches an accuracy of 84.7\% and 65.3\%, respectively, when detecting the emergent intersectional biases unique to African American females and Mexican American females. The probability of random correct attribute detection in EIBD tasks are 9.2\% and 6.1\%. Intersectional biases have the highest magnitude compared to other biases across all language models, potentially disadvantaging members that belong to multiple minority groups in downstream applications.


The current validation set with ground truth information about each word constrains our evaluation to a closed-world machine learning classification task, where we know the category each stimulus belongs to. On the other hand, evaluating the entire semantic space resembles an open-world machine learning problem where millions of stimuli in the entire word embedding vocabulary belong to unknown categories, thus require human-subject annotation studies. In future work, a human subject study can further evaluate the threshold selection criteria, which would require validating a large set of biases retrieved from the entire vocabulary.

 \begin{table*}[t]

  \begin{minipage}[c]{0.68\textwidth}
\centering

\vspace{-3mm}
\label{table:socialbias-measure}
     \resizebox{0.99\textwidth}{!} {%
\begin{tabular}{|p{3mm} l |  r | cc | cc | cc |cc |}
\hline
\multicolumn{3}{| c |}{ \multirow{2}{*}{\textbf{Test}}} &
  \multicolumn{2}{c|}{\textbf{ELMo}} &
  \multicolumn{2}{c|}{\textbf{BERT}} &
  \multicolumn{2}{c|}{\textbf{GPT}} &
  \multicolumn{2}{c |}{\textbf{GPT-2}} \\ \cline{4-11} 
                          \multicolumn{3}{|c|}{}  & \textbf{$d$} & \textbf{$p$}  & \textbf{$d$} & \textbf{$p$}  & \textbf{$d$} & \textbf{$p$} & \textbf{$d$} & \textbf{$p$} \\ \hline

\multirow{2}{*}{\shortstack{C1:}} &  Flowers/Insects  &     random       &  \cellcolor{darkgray}1.40       & $<10^{-30}$           & \cellcolor{darkgray}0.97       &  $<10^{-30}$            & \cellcolor{darkgray}1.04       & $<10^{-30}$          & 0.14       & $<10^{-30}$  \\

  &  Pleasant/Unpleasant$^{\ast}$    &        fixed            & \cellcolor{darkgray}{1.35}                     & $<10^{-30}$                     & \cellcolor{mediumgray}{0.64   }                  & $<10^{-30}$                     & \cellcolor{darkgray}{1.01       }               & $<10^{-30}$                     & \cellcolor{lightgray}{0.21     }                 & $<10^{-30}$                     \\ \hline

\multirow{2}{*}{{\shortstack{C2:}}} & Instruments/Weapons &     random     & \cellcolor{darkgray}1.56       & $<10^{-30}$           & \cellcolor{darkgray}0.94       & $<10^{-30}$  & \cellcolor{darkgray}1.12       & $<10^{-30}$          & \cellcolor{lightgray}-0.27      & $<10^{-30}$      \\
 &  Pleasant/Unpleasant$^{\ast}$    &        fixed           & \cellcolor{darkgray}{1.59}                     & $<10^{-30}$                     & \cellcolor{mediumgray}{0.54}                     & $<10^{-30}$                     & \cellcolor{darkgray}{1.09}                     & $<10^{-30}$                     & \cellcolor{lightgray}{-0.21 }                    & $<10^{-30}$  \\ \hline
 
 \multirow{2}{*}{{\shortstack{C3:}}} & EA/AA names &     random     & \cellcolor{lightgray}0.49       & $<10^{-30}$  & \cellcolor{lightgray}0.44       & $<10^{-30}$   & -0.11      & $<10^{-30}$       & -0.19      & $<10^{-30}$      \\
 &  Pleasant/Unpleasant$^{\ast}$    &        fixed           & \cellcolor{lightgray}{0.47  }                   & $<10^{-30}$                     & \cellcolor{lightgray}{0.31}                     & $<10^{-30}$                     & -0.10                    & $<10^{-30}$                     & 0.09                      & $<10^{-30}$     \\ \hline
 
  \multirow{2}{*}{{\shortstack{C4:}}} & EA/AA names &     random     & 0.15       & $<10^{-30}$  & \cellcolor{lightgray}0.47       & $<10^{-30}$  & 0.01       & $<10^{-2}$       & \cellcolor{lightgray}-0.23      & $<10^{-30}$      \\
 &  Pleasant/Unpleasant$^{\ast}$    &        fixed           & \cellcolor{lightgray}{0.23 }                    & $<10^{-30}$                     & \cellcolor{lightgray}{0.49 }                    & $<10^{-30}$                     & 0.00                     & $0.20$                          & -0.13                     & $<10^{-30}$     \\ \hline
 
  \multirow{2}{*}{{\shortstack{C5:}}} & EA/AA names &     random     & 0.11       & $<10^{-30}$   & 0.02       & $<10^{-7}$        & 0.07       & $<10^{-30}$  & \cellcolor{lightgray}-0.21      & $<10^{-30}$      \\
 &  Pleasant/Unpleasant$^{\ast}$    &        fixed           & 0.17                     & $<10^{-30}$                     & 0.07                     & $<10^{-30}$                     & 0.04                     & $<10^{-27}$                     & -0.01                     & 0.11        \\ \hline 
 
   \multirow{2}{*}{{\shortstack{C6:}}} & Males/Female names &     random     & \cellcolor{darkgray}1.27       & $<10^{-30}$            & \cellcolor{darkgray}0.92       & $<10^{-30}$  & 0.19       & $<10^{-30}$  & \cellcolor{lightgray}0.36       & $<10^{-30}$     \\
 &  Career/Family    &        fixed           & \cellcolor{darkgray}{1.31  }                   & $<10^{-30}$                     & \cellcolor{lightgray}{0.41}                     & $<10^{-30}$                     & 0.11                     & $<10^{-30}$                     & \cellcolor{lightgray}{0.34}                      & $<10^{-30}$        \\ \hline 
 
    \multirow{2}{*}{{\shortstack{C7:}}} & Math/Arts &     random     & \cellcolor{mediumgray}0.64       & $<10^{-30}$   & \cellcolor{lightgray}0.41       & $<10^{-30}$  & \cellcolor{lightgray}0.24       & $<10^{-30}$  & -0.01      & $<10^{-2}$      \\
 &  Male/Female terms    &        fixed           & \cellcolor{darkgray}{0.71 }        &                 $<10^{-30}$              & \cellcolor{lightgray}{0.20  }                   & $<10^{-30}$                     & \cellcolor{lightgray}{0.23}                     & $<10^{-30}$                     & -0.14                     & $<10^{-30}$          \\ \hline 
 
 \multirow{2}{*}{{\shortstack{C8:}}} & Science/Arts &     random     & \cellcolor{lightgray}0.33       & $<10^{-30}$   & -0.07      & $<10^{-30}$        & \cellcolor{lightgray}0.26       & $<10^{-30}$  & -0.16      & $<10^{-30}$     \\
 &  Male/Female terms    &        fixed           & \cellcolor{mediumgray}{0.51     }                & $<10^{-30}$                     & 0.17                     & $<10^{-30}$                     & \cellcolor{lightgray}{0.35}                     & $<10^{-30}$                     & -0.05                     & $<10^{-30}$              \\ \hline 
 
  \multirow{2}{*}{{\shortstack{C9:}}} & Mental/Physical disease &     random     & \cellcolor{darkgray}1.00       & $<10^{-30}$   & \cellcolor{mediumgray}0.53       & $<10^{-30}$  & 0.08      & $<10^{-29}$       & 0.10       & $<10^{-30}$     \\
 &  Temporary/Permanent    &        fixed           & \cellcolor{darkgray}{1.01}                     & $<10^{-30}$                     & \cellcolor{lightgray}{0.40}                     & $<10^{-30}$                     & \cellcolor{lightgray}{-0.23 }                   & $<10^{-30}$                     & \cellcolor{lightgray}{-0.21 }                    & $<10^{-30}$               \\ \hline 
 
   \multirow{2}{*}{{\shortstack{C10:}}} & Young/Old people's names &     random           & 0.11       & $<10^{-30}$  & -0.01       & 0.016        & 0.07       & $<10^{-30}$       & -0.16      & $<10^{-30}$     \\
 &   Pleasant/Unpleasant$^{\ast}$    &        fixed           & \cellcolor{lightgray}{0.24}                     & $<10^{-30}$                     & 0.07                     & $<10^{-30}$                     & 0.04                     & $<10^{-17}$                     & -0.14                     & $<10^{-30}$          \\ \hline 
 
    \multirow{2}{*}{{\shortstack{I1:}}} & AF/EM names &     random           & \cellcolor{darkgray}1.24       & $<10^{-30}$            & \cellcolor{mediumgray}0.77       & $<10^{-30}$   & 0.07       & $<10^{-30}$  & 0.02       & $<10^{-2}$     \\
 &  AF/EM intersectional    &        fixed           & \cellcolor{darkgray}{1.25}                    & $<10^{-30}$                    & \cellcolor{darkgray}{0.98  }                  & $<10^{-30}$                      & \cellcolor{lightgray}{0.23  }                   & $<10^{-30}$                     & -0.19                     & $<10^{-30}$         \\ \hline 
 
     \multirow{2}{*}{{\shortstack{I2:}}} & AF/EM names &     random           & \cellcolor{darkgray}1.25      & $<10^{-30}$            & \cellcolor{mediumgray}0.67      & $<10^{-30}$   & -0.09      & $<10^{-30}$      & 0.02       & $<10^{-2}$    \\
 &  {\small AF emergent/EM intersectional}  &        fixed           & \cellcolor{darkgray}{1.27} & $<10^{-30}$ & \cellcolor{darkgray}{1.00} & $<10^{-30}$                     & \cellcolor{lightgray}{0.23   }                  & $<10^{-30}$                     & -0.14                     & $<10^{-30}$         \\ \hline 
 
      \multirow{2}{*}{{\shortstack{I3:}}} & MF/EM names &     random           & \cellcolor{darkgray}1.31       & $<10^{-30}$            & \cellcolor{mediumgray}0.68       & $<10^{-30}$   & -0.06       & $<10^{-30}$      & \cellcolor{lightgray}0.38       & $<10^{-30}$    \\
 &  MF/EM intersectional    &        fixed           & \cellcolor{darkgray}{1.29}& $<10^{-30}$ & \cellcolor{mediumgray}{0.51} & $<10^{-30}$                     & 0.00                     & 0.81                            & \cellcolor{lightgray}{0.32  }                    & $<10^{-30}$          \\ \hline 
 
       \multirow{2}{*}{{\shortstack{I4:}}} & MF/EM names &     random           & \cellcolor{darkgray} 1.51       & $<10^{-30}$           &\cellcolor{darkgray} 0.86      & $<10^{-30}$  & 0.16       & $<10^{-30}$  & \cellcolor{lightgray}-0.32      & $<10^{-30}$    \\
 & {\small MF emergent/EM intersectional}  &         fixed           & \cellcolor{darkgray}{1.43} & 
 $<10^{-30}$ & \cellcolor{mediumgray}{0.58} & $<10^{-30}$ & \cellcolor{lightgray}{0.20} & $<10^{-30}$ & \cellcolor{lightgray}{-0.25} & $<10^{-30}$        \\ \hline 
\multicolumn{11}{c}{{\small $^{\ast}$Pleasant and unpleasant attributes used to measure valence and attitudes towards targets  from \citet{greenwald1998measuring}.}}\\

\end{tabular}
}
  \end{minipage}\hfill
  \begin{minipage}[c]{0.32\textwidth}
\vspace{-1mm}    \caption{
\textbf{CEAT measures of social and intersectional biases in language models.} We report the overall magnitude of bias in language models with CES ($d$, rounded down) and statistical significance with combined $p$-values ($p$, rounded up). CES pools $N = 10,000$ samples from a random-effects model. The first row for each bias test uses completely random samples, whereas the second row for the bias test uses the same sentences to generate CWE across all neural language models.
    $Ci$ stands for the $i^{th}$ WEAT in \citet{caliskan2017semantics}'s Table 1. $Ii$ stands for our tests constructed for measuring intersectional biases. $A\_$ stands for African Americans, $E\_$ for European Americans, $M\_$ for Mexican Americans, $\_F$ for females, and $\_M$ for males. Light, medium, and dark gray shading of combined $d$ values (CES) indicates small, medium, and large effect size, respectively.  
    } \label{table:socialbias-measure}

  \end{minipage}
\vspace{-3mm}  
  \end{table*}

\subsection{Evaluation of CEAT} Congruent with \citet{caliskan2017semantics}'s WEAT findings, Table~\ref{table:socialbias-measure} presents significant effect sizes for all previously documented and validated biases. GPT-2 exhibited less bias than other neural language models. 
Our method CEAT, designed for CWEs, computes the combined bias score of a distribution of effect sizes present in neural language models. We find that the effect magnitudes of biases reported by Tan and Celis  \citep{tan2019assessing} are individual samples in the distributions generated by CEAT. We can view their method as a special case of CEAT that calculates the individual bias scores of a few pre-selected samples. In order to comprehensively measure the overall bias score in a neural language model, we apply a random-effects model from the meta-analysis literature that computes combined effect size and combined statistical significance from a distribution of bias measurements. As a result, when CEAT reports significant results, some of the corresponding bias scores in prior work are not statistically significant. Furthermore, our results indicate statistically significant bias in the opposite direction in some cases. These negative results suggest that some WEAT stimuli tend to occur in stereotype-incongruent contexts more frequently.

We sampled combinations of CWE $10,000$ times for each CEAT test; nonetheless, we observed varying intensities of the same social bias in different contexts. Using a completely random set vs fixed set of contexts derived from $10,000$ sentences lead to low variance in corresponding bias scores. Using a fixed set of contexts for each model makes it possible to evaluate the magnitude of bias across models for the same variables. Experiments conducted with $1,000$, $5,000$, $10,000$ samples of CWE lead to similar bias scores with low variance. As a result, the number of samples can be adjusted according to computational resources. However, future work on evaluating the lower bound of sampling size with respect to model and corpus characteristics would optimize the sampling process. Accordingly, the computation of overall bias in the language model would become more efficient.

\subsection{IBD, EIBD, and CEAT Results} We report the overall magnitude of bias (CES) and $p$-value  in Table~\ref{table:socialbias-measure}. We pick an example from Table~\ref{table:socialbias-measure} that reflects the great disparity in bias magnitudes between the two models. We present the distribution histograms of effect sizes in Figure~\ref{fig:weat}, which show the overall biases that can be measured with a comprehensive contextualized bias test related to the emergent biases associated with occurrences of stimuli unambiguously regarding Mexican American females (See row I4 in Table~\ref{table:socialbias-measure}) with ELMo and GPT-2. 
The distribution plots for other bias tests are provided in our project repository.



We find that CEAT uncovers more evidence of intersectional bias than gender or racial biases. This findings suggest that, members of multiple minority or disadvantaged groups are associated with the strongest levels of bias in neural language representations. To quantify the intersectional biases in CWEs, we construct tests I1-I4. Tests with Mexican American females tend to have stronger bias with a higher CES than those with African American females. 
Specifically, 13 of 16 instances in intersection-related tests (I1-I4) have significant stereotype-congruent CES; 9 of 12 instances in gender-related tests (C6-C8) have significant stereotype-congruent CES; 8 of 12 instances in  race-related tests (C3-C5) have significant stereotype-congruent CES. In gender bias tests, the gender associations with career and family are stronger than other biased gender associations. In all models, the significantly biased intersectionality associations have larger effect sizes than racial biases.


According to CEAT results in Table~\ref{table:socialbias-measure}, ELMo is the most biased whereas GPT-2 is the least biased with respect to the types of biases CEAT measures. We notice that significant negative CES exist in BERT, GPT and GPT-2, which imply that stereotype-incongruent biases with small effect size exist. 
        
\section{Discussion}
 \label{sec:discussion}

According to our findings, GPT-2 has the highest variance in bias magnitudes followed by GPT, BERT, and ELMo (see an example in Figure~\ref{fig:weat}). The overall magnitude of bias decreases in the same order for the types of biases we measured. The similar number of parameters in these models or the size of the training corpora do not explain the distribution of bias that we observe w.r.t. variance and overall magnitude. However, \citet{ethayarajh2019contextual} note the same descending pattern when measuring words' self-similarity, after adjusting for anisotropy (non-uniform directionality), across their CWE in GPT-2, BERT, and ELMo. (ELMo is compared in three layers due to its architecture.) \citet{ethayarajh2019contextual} also find that upper layers of contextualizing models produce more context-specific representations. Quantifying how contextualized these dynamic embeddings are supports our findings that the highest variance in bias magnitude, low overall bias, and low self-similarity correlate. This correlation may explain the results that we are observing. As more recent models are learning highly-contextualized CWE in upper layers, the representations in highly-contextualized layers are almost overfitting to their contexts. Since words appear in numerous contexts, the more contextualized and diverse a word's representation becomes, the less overall bias and general stereotypical associations.



We present and validate a bias detection method generalizable to identifying biases associated with any social group or intersectional group member. We detect and measure biases associated with Mexican American and African American females in SWE and CWE.
Our emergent intersectional bias measurement results for African American females are in line with previous findings \citep{may2019measuring,tan2019assessing}.
IBD and EIBD can detect intersectional biases from SWE with high accuracy in an unsupervised manner by following a lexicon induction strategy \cite{hatzivassiloglou1997predicting}. This approach can be complementary to the stimuli list predefined by social psychologists.
Our current intersectional bias detection validation approach can be used to identify association thresholds when generalizing this work to the entire word embedding dictionary. Exploring all the potential biases associated with targets is left to future work since it requires extensive human subject validation studies in collaboration with social psychologists. We list all the stimuli representing biased associations in the supplementary materials. To name a few, the superset of intersectional biases associated with African American females are: aggressive, assertive, athletic, bigbutt, confident, darkskinned, fried-chicken, ghetto, loud, overweight, promiscuous, unfeminine, unintelligent, unrefined. Emergent intersectional biases associated with African American females are: aggressive, assertive, bigbutt, confident, darkskinned, fried-chicken, overweight, promiscuous, unfeminine. The superset of intersectional biases associated with Mexican American females are: attractive, cook, curvy, darkskinned, feisty, hardworker, loud, maids, promiscuous, sexy, short, uneducated, unintelligent. Emergent intersectional biases associated with Mexican American females are:  cook, curvy, feisty, maids, promiscuous, sexy.

We follow the conventional method of using the most frequent given names in a social group that signal group membership in order to accurately represent targets \citep{caliskan2017semantics,greenwald1998measuring}. 
Our results indicate that the conventional method that relies on stimuli selected by experts in social psychology works accurately. Prior work on lexicon induction methods compensates for the lack of existing annotated data on valence \cite{hatzivassiloglou1997predicting, riloff2003learning, turney2003measuring}. Nevertheless, principled and robust lexicon induction methods that can be validated in this domain, when measuring the representation accuracy of target group lexica or any semantic concept. Developing these principled methods is left to future work.  

Semantics of languages can be represented by the distributional statistics of word co-occurrences \cite{firth1957synopsis, harris1954distributional}. Consequently, our methods are language agnostic and can be applied to neural language models as well as word embeddings in any language as long as the stimuli for accurately representing the semantics of concepts are available. Project Implicit (\url{https://implicit.harvard.edu/implicit}) has been hosting IATs for human subjects all over the world in numerous languages for two decades. As a result, their IATs, that inspired WEATs, provide stimuli for targets and attributes in numerous languages. We leave generalizing our methods to other languages to future work since state-of-the-art neural language models are not widely or freely available for languages other than English as of 2021.


 When simulating contexts for WEAT, we make an assumption that the Reddit corpus represents naturally occurring sentences. Nevertheless, we acknowledge that the Reddit corpus also reflects the biases of the underlying population contributing to its corpus. Studying the accuracy of simulating the most common distribution of contexts and co-occurring stimuli is left to future work since we don't have validated ground truth data for evaluating the distribution parameters of contexts in large-scale corpora. Instead, for evaluation, validation, and comparison, we rely on validated ground truth information about biases documented by \citet{caliskan2017semantics} in word embeddings as well as biases documented by millions of people over decades via the implicit association literature \cite{nosek2002harvesting} and \citet{ghavami2013intersectional}'s intersectional biases.

Given the energy and funding considerations, we are not able to train these language models on the same large-scale corpora to compare how a neural language model's architecture learns biases, because the training processes for these models are computationally and financially expensive \cite{bender2021dangers}. The size of state-of-the-art models increase by at least a factor of 10 every year. BERT-Large from 2018 has 355 million parameters, GPT-2 from early 2019 reaches 1.5 billion, and GPT-3 from mid-2020 finally gets to 175 billion parameters. The GPT-2 model used 256 Google Cloud TPU v3 cores for training, which costs 256 US dollars per hour. GPT-2 requires approximately 168 hours or 1 week of training on 32 TPU v3 chips \cite{strubell2019energy}. GPT-3 is estimated to cost $\sim$12 million US dollars \cite{floridi2020gpt} and we are not able to get access to its embeddings or training corpora. Regardless, measuring the scope of biases with validated bias quantification and meta-analysis methods, we are able to compare the biased associations learned by neural language models that are widely used. Being able to study neural language models comprehensively is critical since they are replacing SWE in many NLP applications due to their high accuracy in various machine learning tasks.


We would like to conclude the discussion with our ethical concerns regarding the dual use of IBD and EIBD, that can detect stereotypical associations for an intersectional group or disadvantaged individuals. Words retrieved by our methods may be used in the generation of offensive or stereotypical content that perpetuates or amplifies existing biases. For example, information influence operations in the 1970s used \citet{osgood1964semantic}'s semantic differential technique among human subjects to retrieve the words that would most effectively induce a negative attitude in a South American population towards their administration \cite{landis1982cia}. Similarly, biased neural language models may be exploited to automate large-scale information influence operations that intend to sow discord among social groups \citet{toney2020pro, toney2020valnorm}. The biased outputs of these language models, that get recycled in future model generation's training corpora, may lead to an AI bias feedback cycle.

\section{Conclusion}
\label{sec:conclusion}

We introduce methods called IBD and EIBD to identify biases associated with members of multiple minority groups. These methods automatically detect the intersectional biases and emergent intersectional biases captured by word embeddings. Intersectional biases associated with African American and Mexican American females have the highest effect size compared to other social biases. Complementary to pre-defined sets of attributes to measure widely known biases, our methods automatically discover biases.
IBD reaches an accuracy of 81.6\% and 82.7\% in detection, respectively, when validating on the intersectional biases of African American females and Mexican American females.
EIBD reaches an accuracy of 84.7\%  and 65.3\%  in  detection, respectively, when validating on the emergent intersectional biases of African American females and Mexican American females.

We present CEAT to measure biases identified by IBD and EIBD in language models. CEAT uses a random-effects model to comprehensively measure social biases embedded in neural language models that contain a distribution of context-dependent biases. CEAT simulates this distribution by sampling ($N=10,000$) combinations of CWEs without replacement from a large-scale natural language corpus. 
Unlike prior work that focuses on a limited number of contexts defined by templates to measure the magnitude of particular biases, CEAT provides a comprehensive measurement of overall bias in contextualizing language models. Our results indicate that ELMo is the most biased, followed by BERT, and GPT. GPT-2 is the least biased language model with respect to the social biases we investigate. The overall magnitude of bias negatively correlates with the level of contextualization in the language model. Understanding how the architecture of a language model contributes to biased and contextualized word representations can help mitigate the harmful effects to society in downstream applications.

\newpage
\section{Appendices}
\subsection{Formal Definition of WEAT}
We present a formal definition of \citet{caliskan2017semantics}'s WEAT. Let $X$ and $Y$ be two sets of target words of equal size, and $A$, $B$ be two sets of attribute words. Let $cos(\vec{a},\vec{b})$ stand for the cosine similarity between the embeddings of words $a$ and $b$. Here, the vector $\vec{a}$ is the embedding for word $a$. The test statistic is 
\[ s(X,Y,A,B) = \sum_{x\in X}{s(x,A,B)} -  \sum_{y\in Y}{s(y,A,B)} \]
where 
\[ s(w,A,B) = mean_{a \in A}cos(\vec{w}, \vec{a})-mean_{b \in B}cos(\vec{w}, \vec{b}) \]

A permutation test calculates the statistical significance of association $s(X,Y,A,B)$.  The one-sided $p-value$ is 
\[ P = Pr_{i} [s(X_{i},Y_{i},A,B)>s(X,Y,A,B))] \]
where $\{(X_i,Y_i)\}_{i}$ represents all the partitions of $X\cup Y$ in two sets of equal size. Random permutations of these stimuli sets represent the null hypothesis  as if the biased associations did not exist so that we can perform a statistical significance test by measuring the unlikelihood of the null hypothesis, given the effect size of WEAT.

The effect size of bias is calculated as 
\[ ES = \frac{mean_{x \in X}s(x,A,B)-mean_{y \in Y}s(y,A,B)}{std\_dev_{w \in X\bigcup Y}s(w,A,B)} \]

\subsection{Formal Definition of EIBD}
We first detect $C_{11}$'s intersectional biases $W_{IB}$ with IBD.
Then, we detect the biased attributes associated with only one constituent category of the intersectional group $C_{11}$ (e.g., associated only with race $S_{1n}$ -  or only with gender $S_{m1}$). Each intersectional category $C_{1n}$ has M constituent subcategories $S_{in},i=1,...M$ and category $C_{m1}$ has N constituent subcategories $S_{mj},j=1,...,N$.
$S_{1n}$ and $S_{m1}$ are the constituent subcategories of intersectional group $C_{11}$.

There are in total $M+N$ groups defined by all the single constituent subcategories. We use all $M+N$ groups  to build WEFAT pairs $P_i = (S_{1n},S_{in}),i=1,...,M$ and $P_j=(S_{m1},S_{mj}),j=1,...N$. Then, we detect lists of words associated with each pair $W_i,i=1,...M$ and $W_j,j=1,...,N$ based on the same positive threshold $t_{mn}$ used in IBD. We detect the attributes highly associated with the constituent subcategories $S_{1n}$ and $S_{m1}$ of the target intersectional group $C_{11}$ from all $(M+N)$ WEFAT pairs. We define the words associated with emergent intersectional biases of group $C_{11}$ as $W_{EIB}$ and these words are identified by the formula
\vspace{-3mm}
\[  W_{EIB} = (\bigcup_{i=1}^{M} (W_{IB}-W_{i}))
\bigcup (\bigcup_{j=1}^{N} (W_{IB}-W_{j})) \]
\noindent where 
\vspace{-6mm}
\[  W_i = \{w|s(w,S_{1n},S_{in})>t_{mn}, w \in W_{IB}\}\] 

\noindent and 
\vspace{-6mm}
\[ W_j= \{w|s(w,S_{m1},S_{mj})>t_{mn}, w \in W_{IB}\}\]





\subsection{Random-Effects Model Details}
Each effect size is calculated by 
\[ ES_{i} = \frac{mean_{x \in X}s(x,A,B)-mean_{y \in Y}s(y,A,B)}{std\_dev_{w \in X\bigcup Y}s(w,A,B)} \]

 The estimation of in-sample variance is $V_{i}$, which is the square of $std\_dev_{w \in X\bigcup Y}s(w,A,B)$. 
 We use the same principle as estimation of the variance components in ANOVA to measure the between-sample variance $\sigma^{2}_{between}$, which is calculated as:
\[\sigma^{2}_{between}=\left\{
\begin{aligned}
    &\frac{Q-(N-1)}{c} & if \hspace{2mm}\*\* Q \geq N-1\\
    &0 & if\hspace{2mm}\*\* Q < N-1
\end{aligned}
\right.
\]
where 
\vspace{-3mm}
\[
W_{i} = \frac{1}{V_{i}}
\]

\vspace{-3mm}

\[c = \sum W_{i} - \frac{\sum W_{i}^{2}}{\sum W_{i}} \hspace{2mm}  \& \hspace{2mm} Q = \sum W_{i} ES_{i}^{2} - \frac{(\sum W_{i}ES_{i})^2}{\sum W_{i}} \]

The weight $v_{i}$ assigned to each WEAT is the inverse of the sum of estimated in-sample variance $V_{i}$ and estimated between-sample variance in the distribution of random-effects $\sigma^{2}_{between}$.
\[
v_{i} = \frac{1}{V_{i} + \sigma^{2}_{between}}
\]

CES, which is the sum of the weighted effect sizes divided by the sum of all weights, is then computed as
\[
CES = \frac{\sum_{i=1}^{N}v_{i}ES_{i}}{\sum_{i=1}^{N}v_{i}}
\]

To derive the hypothesis test, we calculate the standard error (SE) of CES as the square root of the inverse of the sum of the weights.
\[
SE(CES) = \sqrt{\frac{1}{\sum_{i=1}^{N}v_{i}}}
\]
Based on the central limit theorem, the limiting form of the distribution of $\frac{CES}{SE(CES)}$ is the standard normal distribution \cite{montgomery2010applied}.
Since we notice that some CES are negative, we use a two-tailed $p-value$ which can test the significance of biased associations in two directions.
The two-tailed $p-value$ of the hypothesis that there is no difference between all the contextualized variations of the two sets of target words in terms of their relative similarity to two sets of attribute words is given by the following formula,
 where $\Phi$ is the standard normal cumulative distribution function and $SE$ stands for the standard error.
 \[ P_{combined}(X,Y,A,B) = 2 \times [1 - \Phi ( | \frac{CES}{SE(CES)} | ) ] \]



\subsection{Meta-Analysis Details for CEAT}

In this section, we first construct all  CEAT in the main paper (C1-C10,I1-I4) with sample size $N=1,000$ to provide a comparison of results with different sample sizes. We report CES $d$ and combined $p-value$ $p$ in Table~\ref{table:supp-main}. We replicate these results with $N=1,000$ instead of using the original $N=10,000$ to show that even with $N=1,000$, we get valid results. Accordingly, we proceed to calculate all types of biases associated with intersectional groups based on the attributes used in original WEAT.  
We notice that there are five tests which are significant with sample size $N=10,000$ but insignificant with sample size $N=1,000$. They are  C10 with Bert,  C4 with GPT, C7 with GPT-2, I3 with GPT-2 and I4 with GPT-2. We also notice that CES of same test can be different with different sample size but all differences are smaller than $0.1$.


\begin{table*}[t]
\caption{\textbf{CEAT from main paper (C1-C10,I1-I4) with sample size $N=1,000$ as opposed to the $N=10,000$ hyper-parameter in the main paper.} We report the CES ($d$) and combined $p-values$ of all  CEAT  ($p$) in the main paper with sample size $N=1,000$. We observe that all of the results are consistent with the CES and $p-values$ reported in the main paper on Table 1. Light, medium, and dark gray shading of combined $d$ values (CES) indicates small, medium, and large effect size, respectively. There are five tests which are significant with sample size $N=10,000$ but not significant with sample size $N=1,000$. However, these have small effect sizes and as a result we don't expect statistical significance. According to our experiments, the Spearman correlation between WEAT's effect size and $p-value$ is $\rho=0.99$. Smaller effect sizes are expected to have insignificant p-values. Accordingly, all of the results under $N=1,000$ are consistent with the main findings. The notable yet consistent differences are C10 with Bert, C4 with GPT, C7 with GPT-2, I3 with GPT-2, and I4 with GPT-2.  CES varies minimally with different sample size ($N$), but the differences of the results are smaller than $0.1$, suggesting the degree of effect size remains consistent. In edge cases, where statistical significance or effect size is close to a significance threshold, gradually increasing $N$, in increments of $N=+500$  would provide more reliable results. $A\_$ stands for African Americans. $E\_$ stands for European Americans. $M\_$ stands for Mexican Americans. $\_F$ stands for females. $\_M$ stands for males.\\}
\label{table:supp-main}
  \resizebox{\textwidth}{!}{%
\begin{tabular}{@{}lcccccccc@{}}
\toprule
\textbf{Test} &
  \multicolumn{2}{c}{\textbf{ELMo}} &
  \multicolumn{2}{c}{\textbf{BERT}} &
  \multicolumn{2}{c}{\textbf{GPT}} &
  \multicolumn{2}{c}{\textbf{GPT-2}} \\ \cmidrule(l){2-9} 
                                         & $d$  & $p$         & $d$   & $p$         & $d$   & $p$         & $d$   & $p$         \\ \midrule
C1: Flowers/Insects, P/U$^{\ast}$ - Attitude                 & \cellcolor{darkgray}1.39 & $<10^{-30}$ & \cellcolor{darkgray}0.96  & $<10^{-30}$    & \cellcolor{darkgray}1.05  & $<10^{-30}$    & 0.13  & $<10^{-30}$    \\
C2: Instruments/Weapons, P/U$^{\ast}$ - Attitude             & \cellcolor{darkgray}1.56 & $<10^{-30}$ & \cellcolor{darkgray}0.93  & $<10^{-30}$    & \cellcolor{darkgray}1.13  & $<10^{-30}$    & \cellcolor{lightgray}-0.28 & $<10^{-30}$    \\
C3: EA/AA names, P/U$^{\ast}$ - Attitude                     & \cellcolor{lightgray}0.48 & $<10^{-30}$ &\cellcolor{lightgray} 0.45  & $<10^{-30}$ & -0.11 & $<10^{-30}$ & \cellcolor{lightgray}-0.20 & $<10^{-30}$ \\
C4: EA/AA names, P/U$^{\ast}$ - Attitude                     & 0.16 & $<10^{-30}$ & \cellcolor{lightgray}0.49  & $<10^{-30}$ & 0.00  & 0.70        & \cellcolor{lightgray}-0.23 & $<10^{-30}$ \\
C5: EA/AA names, P/U$^{\ast}$ - Attitude                     & 0.12 & $<10^{-30}$ & 0.04  & $<10^{-2}$  & 0.05  & $<10^{-4}$  & -0.17 & $<10^{-30}$ \\
C6: Males/Female names, Career/Family    & \cellcolor{darkgray}1.28 & $<10^{-30}$ & \cellcolor{darkgray}0.91  & $<10^{-30}$ & \cellcolor{lightgray}0.21  & $<10^{-30}$ & \cellcolor{lightgray}0.34  & $<10^{-30}$ \\
C7: Math/Arts, Male/Female terms         & \cellcolor{mediumgray}0.65 & $<10^{-30}$ & \cellcolor{lightgray}0.42  & $<10^{-30}$ & \cellcolor{lightgray}0.23  & $<10^{-30}$ & 0.00  & 0.81        \\
C8: Science/Arts, Male/Female terms      & \cellcolor{lightgray}0.32 & $<10^{-30}$ & -0.07 & $<10^{-4}$  & \cellcolor{lightgray}0.26  & $<10^{-30}$ & -0.16 & $<10^{-30}$ \\
C9: Mental/Physical disease, Temporary/Permanent              & \cellcolor{darkgray}0.99 & $<10^{-30}$ & \cellcolor{mediumgray}0.55  & $<10^{-30}$ & 0.07  & $<10^{-2}$  & 0.04  & 0.04        \\
C10: Young/Old people's names, P/U$^{\ast}$ - Attitude       & 0.11 & $<10^{-19}$ & 0.00  & 0.90        & 0.04  & $<10^{-2}$  & -0.17 & $<10^{-30}$ \\
I1: AF/EM, AF/EM intersectional          & \cellcolor{darkgray}1.24 & $<10^{-30}$ & \cellcolor{mediumgray}0.76  & $<10^{-30}$ & 0.05  & $<10^{-3}$  & 0.05  & 0.06        \\
I2: AF/EM, AF emergent/EM intersectional & \cellcolor{darkgray}1.24 & $<10^{-30}$ & \cellcolor{mediumgray}0.70  & $<10^{-30}$ & -0.12 & $<10^{-30}$ & 0.03  & 0.26        \\
I3: MF/EM, MF/EM intersectional          & \cellcolor{darkgray}1.30 & $<10^{-30}$ & \cellcolor{mediumgray}0.69  & $<10^{-30}$ & -0.08 & $<10^{-30}$ & \cellcolor{lightgray}0.36  & $<10^{-30}$ \\
I4: MF/EM, MF emergent/EM intersectional &
  \cellcolor{darkgray}1.52 &
  $<10^{-30}$ &
  \cellcolor{darkgray}0.87 &
  $<10^{-30}$ &
  0.14 &
  $<10^{-27}$ &
  \cellcolor{lightgray}-0.26 &
  $<10^{-30}$ \\ \bottomrule
  \multicolumn{9}{c}{$^{\ast}$Unpleasant and pleasant attributes used to measure valence and attitudes towards targets \cite{greenwald1998measuring}.}
\end{tabular}}
\end{table*}

We also construct four types of  supplementary  CEAT  for all pairwise combinations of six intersectional groups: African American females (AF), African American males (AM), Mexican American females (MF), Mexican American males (MM), European American females (EF), European American males (EM). We use two intersectional groups as two target social groups. For each pairwise combination, we build four  CEAT : first, measure attitudes with words representing pleasantness and unpleasantness as two attribute groups (as in C1); second, measure career and family associations that are particularly important in gender stereotypes with the corresponding two attribute groups (as in C6); third, similar to the career-family stereotypes for gender, measure math and arts associations that are particularly important in gender stereotypes with the corresponding two attribute groups (as in C7); fourth, similar to the math-arts stereotypes for gender, measure science (STEM) and arts associations that are particularly important in gender stereotypes with the corresponding two attribute groups (as in C8). We report the CES ($d$) and combined $p-values$ ($p$) in Table 2 with sample size $N=1,000$. All of these attributes are from the C1, C6, C7 and C8 WEAT of Caliskan et al. \cite{caliskan2017semantics}.


\begin{table*}
\begin{center}
\captionof{table}{\textbf{CEAT for intersectional groups with sample size $N=1,000$.} We construct 4 types of new CEAT with all pairwise combinations of intersectional groups. We use two intersectional groups as two target social groups. We use 1) pleasant/unpleasant 2) career/family 3) math/arts 4) science/arts as two attribute groups. 
We report the CES $d$ and combined $p-value$ $p$. Light, medium, and dark gray shading of combined $d$ values (CES) indicates small, medium, and large effect size respectively. $A\_$ stands for African Americans. $E\_$ stands for European Americans. $M\_$ stands for Mexican Americans. $\_F$ stands for females. $\_M$ stands for males.}

    \resizebox{0.65\textwidth}{!}{%
 \begin{small}

\begin{tabular}[h]{@{}l c c c c c c c c @{}}
\label{table:supp-new}\\
\textbf{Test} &  
\multicolumn{2}{c}{\textbf{ELMo}} &
\multicolumn{2}{c}{\textbf{BERT}} &
\multicolumn{2}{c}{\textbf{GPT}} &
\multicolumn{2}{c}{\textbf{GPT-2} } \\ \cmidrule(l){2-9} 
 & $d$   & $p$         & $d$    & $p$         & $d$   & $p$         & $d$   & $p$ \\  \midrule
\hline

EM/EF, P/U$^{\ast}$ - Attitude           & \cellcolor{lightgray}-0.49 & $<10^{-30}$ & \cellcolor{lightgray}-0.33 & $<10^{-30}$ & -0.01 & 0.60        & \cellcolor{mediumgray}-0.53 & $<10^{-30}$ \\
EM/EF, Career/Family & \cellcolor{darkgray}1.15  & $<10^{-30}$ & \cellcolor{mediumgray}0.73  & $<10^{-30}$ & \cellcolor{lightgray}0.34  & $<10^{-30}$ & \cellcolor{lightgray}0.41  & $<10^{-30}$ \\
EM/EF, Math/Arts     & \cellcolor{lightgray}0.44  & $<10^{-30}$ & \cellcolor{lightgray}0.34  & $<10^{-30}$ & 0.13  & $<10^{-25}$ & \cellcolor{lightgray}-0.41 & $<10^{-30}$ \\
EM/EF, Science/Arts  & \cellcolor{lightgray}0.37  & $<10^{-30}$ & -0.11 & $<10^{-30}$ & 0.07  & $<10^{-6}$  & -0.04 & 0.02        \\
EM/AM, P/U$^{\ast}$ - Attitude           & \cellcolor{mediumgray}0.57  & $<10^{-30}$ & \cellcolor{lightgray}0.40  & $<10^{-30}$ & 0.04  & $<10^{-2}$  & \cellcolor{lightgray}-0.34 & $<10^{-30}$ \\
EM/AM, Career/Family & \cellcolor{lightgray}0.32  & $<10^{-30}$ & 0.16  & $<10^{-30}$ & \cellcolor{lightgray}-0.36 & $<10^{-30}$ & \cellcolor{lightgray}0.42  & $<10^{-30}$ \\
EM/AM, Math/Arts     & \cellcolor{lightgray}-0.28 & $<10^{-30}$ & -0.04 & $<10^{-2}$  & -0.05 & $<10^{-30}$ & \cellcolor{lightgray}-0.45 & $<10^{-30}$ \\
EM/AM, Science/Arts  & 0.02  & 0.10        & -0.18 & $<10^{-30}$ & 0.17  & $<10^{-30}$ & \cellcolor{lightgray}-0.20 & $<10^{-30}$ \\
EM/AF, P/U$^{\ast}$ - Attitude           & \cellcolor{lightgray}-0.35 & $<10^{-30}$ & 0.10  & $<10^{-11}$ & -0.12 & $<10^{-30}$ & \cellcolor{mediumgray}-0.60 & $<10^{-30}$ \\
EM/AF, Career/Family & \cellcolor{darkgray}1.10  & $<10^{-30}$ & \cellcolor{darkgray}0.90  & $<10^{-30}$ & \cellcolor{lightgray}0.20  & $<10^{-30}$ & \cellcolor{mediumgray}0.62  & $<10^{-30}$ \\
EM/AF, Math/Arts     & 0.11  & $<10^{-19}$ & \cellcolor{mediumgray}0.72  & $<10^{-30}$ & 0.14  & $<10^{-23}$ & \cellcolor{mediumgray}-0.62 & $<10^{-30}$ \\
EM/AF, Science/Arts  & \cellcolor{mediumgray}0.56  & $<10^{-30}$ & \cellcolor{lightgray}0.29  & $<10^{-30}$ & \cellcolor{lightgray}0.24  & $<10^{-30}$ & -0.19 & $<10^{-30}$ \\
EM/MM, P/U$^{\ast}$ - Attitude           & -0.15 & $<10^{-30}$ & \cellcolor{lightgray}0.42  & $<10^{-30}$ & -0.17 & $<10^{-30}$ & \cellcolor{lightgray}-0.20 & $<10^{-30}$ \\
EM/MM, Career/Family & 0.01  & 0.46        & \cellcolor{lightgray}0.28  & $<10^{-30}$ & \cellcolor{lightgray}-0.32 & $<10^{-30}$ & \cellcolor{lightgray}0.33  & $<10^{-30}$ \\
EM/MM, Math/Arts     & 0.06  & $<10^{-5}$  & \cellcolor{lightgray}-0.22 & $<10^{-30}$ & \cellcolor{lightgray}0.45  & $<10^{-30}$ & \cellcolor{lightgray}-0.38 & $<10^{-30}$ \\
EM/MM, Science/Arts  & \cellcolor{lightgray}0.21  & $<10^{-30}$ & \cellcolor{lightgray}-0.27 & $<10^{-30}$ & \cellcolor{mediumgray}0.62  & $<10^{-30}$ & \cellcolor{lightgray}-0.37 & $<10^{-30}$ \\
EM/MF, P/U$^{\ast}$ - Attitude           & \cellcolor{darkgray}-0.82 & $<10^{-30}$ & -0.19 & $<10^{-30}$ & \cellcolor{lightgray}-0.34 & $<10^{-30}$ & \cellcolor{mediumgray}-0.60 & $<10^{-30}$ \\
EM/MF, Career/Family & \cellcolor{darkgray}1.14  & $<10^{-30}$ & \cellcolor{mediumgray}0.68  & $<10^{-30}$ & 0.09  & $<10^{-11}$ & \cellcolor{mediumgray}0.68  & $<10^{-30}$ \\
EM/MF,Math/Arts      & \cellcolor{mediumgray}0.69  & $<10^{-30}$ & \cellcolor{lightgray}0.27  & $<10^{-30}$ & \cellcolor{lightgray}0.28  & $<10^{-30}$ & \cellcolor{mediumgray}-0.78 & $<10^{-30}$ \\
EM/MF, Science/Arts  & \cellcolor{lightgray}0.33  & $<10^{-30}$ & 0.11  & $<10^{-13}$ & \cellcolor{lightgray}0.41  & $<10^{-30}$ & \cellcolor{lightgray}-0.29 & $<10^{-30}$ \\
EF/AM, P/U$^{\ast}$ - Attitude           & \cellcolor{darkgray}0.95  & $<10^{-30}$ & \cellcolor{mediumgray}0.70  & $<10^{-30}$ & 0.06  & $<10^{-5}$  & 0.09  & $<10^{-17}$ \\
EF/AM, Career/Family & \cellcolor{darkgray}-0.98 & $<10^{-30}$ & \cellcolor{mediumgray}-0.62 & $<10^{-30}$ & \cellcolor{mediumgray}-0.63 & $<10^{-30}$ & 0.11  & $<10^{-21}$ \\
EF/AM, Math/Arts     & \cellcolor{mediumgray}-0.66 & $<10^{-30}$ & \cellcolor{lightgray}-0.41 & $<10^{-30}$ & -0.15 & $<10^{-30}$ & -0.10 & $<10^{-30}$ \\
EF/AM, Science/Arts  & \cellcolor{lightgray}-0.30 & $<10^{-30}$ & -0.08 & $<10^{-30}$ & 0.11  & $<10^{-13}$ & -0.19 & $<10^{-30}$ \\
EF/AF, P/U$^{\ast}$ - Attitude           & 0.09  & $<10^{-22}$ & \cellcolor{mediumgray}0.50  & $<10^{-30}$ & -0.15 & $<10^{-30}$ & \cellcolor{lightgray}-0.20 & $<10^{-30}$ \\
EF/AF, Career/Family & 0.04  & $<10^{-7}$  & \cellcolor{lightgray}0.22  & $<10^{-30}$ & -0.16 & $<10^{-30}$ & \cellcolor{lightgray}0.33  & $<10^{-30}$ \\
EF/AF, Math/Arts     & \cellcolor{lightgray}-0.33 & $<10^{-30}$ & \cellcolor{lightgray}0.39  & $<10^{-30}$ & -0.01 & 0.44        & \cellcolor{lightgray}-0.35 & $<10^{-30}$ \\
EF/AF, Science/Arts  & \cellcolor{lightgray}0.23  & $<10^{-30}$ & \cellcolor{lightgray}0.43  & $<10^{-30}$ & 0.18  & $<10^{-30}$ & \cellcolor{lightgray}-0.20 & $<10^{-30}$ \\
EF/MM, P/U$^{\ast}$ - Attitude           & \cellcolor{lightgray}0.38  & $<10^{-30}$ & \cellcolor{mediumgray}0.70  & $<10^{-30}$ & -0.19 & $<10^{-30}$ & \cellcolor{lightgray}0.32  & $<10^{-30}$ \\
EF/MM, Career/Family & \cellcolor{darkgray}-1.10 & $<10^{-30}$ & \cellcolor{lightgray}-0.45 & $<10^{-30}$ & \cellcolor{mediumgray}-0.65 & $<10^{-30}$ & -0.02 & 0.14        \\
EF/MM, Math/Arts     & \cellcolor{lightgray}-0.34 & $<10^{-30}$ & \cellcolor{mediumgray}-0.55 & $<10^{-30}$ & \cellcolor{lightgray}0.37  & $<10^{-30}$ & -0.02 & 0.28        \\
EF/MM, Science/Arts  & -0.18 & $<10^{-30}$ & \cellcolor{lightgray}-0.21 & $<10^{-30}$ & \cellcolor{mediumgray}0.54  & $<10^{-30}$ & \cellcolor{lightgray}-0.36 & $<10^{-30}$ \\
EF/MF, P/U$^{\ast}$ - Attitude           & \cellcolor{lightgray}-0.42 & $<10^{-30}$ & 0.19  & $<10^{-30}$ & \cellcolor{lightgray}-0.33 & $<10^{-30}$ & -0.15 & $<10^{-30}$ \\
EF/MF, Career/Family & -0.09 & $<10^{-30}$ & -0.07 & $<10^{-30}$ & \cellcolor{lightgray}-0.23 & $<10^{-30}$ & \cellcolor{lightgray}0.43  & $<10^{-30}$ \\
EF/MF, Math/Arts     & \cellcolor{lightgray}0.30  & $<10^{-30}$ & -0.05 & $<10^{-30}$ & 0.17  & $<10^{-30}$ & -0.55 & $<10^{-30}$ \\
EF/MF, Science/Arts  & -0.01 & 0.40        & \cellcolor{lightgray}0.25  & $<10^{-30}$ & \cellcolor{lightgray}0.37  & $<10^{-30}$ & \cellcolor{lightgray}-0.30 & $<10^{-30}$ \\
AM/AF, P/U$^{\ast}$ - Attitude           & \cellcolor{mediumgray}-0.79 & $<10^{-30}$ & \cellcolor{lightgray}-0.32 & $<10^{-30}$ & -0.19 & $<10^{-30}$ & \cellcolor{lightgray}-0.24 & $<10^{-30}$ \\
AM/AF, Career/Family & \cellcolor{darkgray}0.94  & $<10^{-30}$ & \cellcolor{darkgray}0.84  & $<10^{-30}$ & \cellcolor{mediumgray}0.50  & $<10^{-30}$ & 0.17  & $<10^{-30}$ \\
AM/AF, Math/Arts     & \cellcolor{lightgray}0.34  & $<10^{-30}$ & \cellcolor{mediumgray}0.79  & $<10^{-30}$ & 0.16  & $<10^{-30}$ & -0.17 & $<10^{-30}$ \\
AM/AF, Science/Arts  & \cellcolor{mediumgray}0.50  & $<10^{-30}$ & \cellcolor{lightgray}0.47  & $<10^{-30}$ & 0.07  & $<10^{-7}$  & -0.02 & 0.15        \\
AM/MM, P/U$^{\ast}$ - Attitude           & \cellcolor{mediumgray}-0.72 & $<10^{-30}$ & 0.02  & 0.10        & \cellcolor{lightgray}-0.20 & $<10^{-30}$ & \cellcolor{lightgray}0.20  & $<10^{-30}$ \\
AM/MM, Career/Family & \cellcolor{lightgray}-0.28 & $<10^{-30}$ & 0.16  & $<10^{-30}$ & 0.07  & $<10^{-7}$  & -0.12 & $<10^{-30}$ \\
AM/MM, Math/Arts     & \cellcolor{lightgray}0.33  & $<10^{-30}$ & -0.16 & $<10^{-30}$ & \cellcolor{mediumgray}0.51  & $<10^{-30}$ & 0.08  & $<10^{-9}$  \\
AM/MM, Science/Arts  & 0.13  & $<10^{-30}$ & -0.13 & $<10^{-30}$ & \cellcolor{lightgray}0.45  & $<10^{-30}$ & -0.16 & $<10^{-30}$ \\
AM/MF, P/U$^{\ast}$ - Attitude           & \cellcolor{darkgray}-1.15 & $<10^{-30}$ & \cellcolor{mediumgray}-0.57 & $<10^{-30}$ & \cellcolor{lightgray}-0.38 & $<10^{-30}$ & \cellcolor{lightgray}-0.22 & $<10^{-30}$ \\
AM/MF, Career/Family &  \cellcolor{darkgray}0.96  & $<10^{-30}$ & \cellcolor{mediumgray}0.56  & $<10^{-30}$ & \cellcolor{lightgray}0.41  & $<10^{-30}$ & \cellcolor{lightgray}0.27  & $<10^{-30}$ \\
AM/MF,  Math/Arts    & \cellcolor{darkgray}0.87  & $<10^{-30}$ & \cellcolor{lightgray}0.36  & $<10^{-30}$ & \cellcolor{lightgray}0.31  & $<10^{-30}$ & \cellcolor{lightgray}-0.38 & $<10^{-30}$ \\
AM/MF, Science/Arts  & \cellcolor{lightgray}0.30  & $<10^{-30}$ & \cellcolor{lightgray}0.30  & $<10^{-30}$ & \cellcolor{lightgray}0.27  & $<10^{-30}$ & -0.14 & $<10^{-30}$ \\
AF/MM, P/U$^{\ast}$ - Attitude           & \cellcolor{lightgray}0.26  & $<10^{-30}$ & \cellcolor{lightgray}0.33  & $<10^{-30}$ & -0.04 & $<10^{-30}$ & \cellcolor{lightgray}0.46  & $<10^{-30}$ \\
AF/MM, Career/Family & \cellcolor{darkgray}-1.07 & $<10^{-30}$ & \cellcolor{mediumgray}-0.64 & $<10^{-30}$ & \cellcolor{mediumgray}-0.54 & $<10^{-30}$ & \cellcolor{lightgray}-0.31 & $<10^{-30}$ \\
AF/MM, Math/Arts     & -0.03 & 0.03        & \cellcolor{darkgray}-0.90 & $<10^{-30}$ & \cellcolor{lightgray}0.37  & $<10^{-30}$ & \cellcolor{lightgray}0.29  & $<10^{-30}$ \\
AF/MM, Science/Arts  & \cellcolor{lightgray}-0.38 & $<10^{-30}$ & \cellcolor{mediumgray}-0.56 & $<10^{-30}$ & \cellcolor{lightgray}0.43  & $<10^{-30}$ & -0.18 & $<10^{-30}$ \\
AF/MF, P/U$^{\ast}$ - Attitude           & \cellcolor{lightgray}-0.43 & $<10^{-30}$ & \cellcolor{lightgray}-0.33 & $<10^{-30}$ & -0.19 & $<10^{-30}$ & -0.01 & 0.48        \\
AF/MF, Career/Family & -0.15 & $<10^{-30}$ & \cellcolor{lightgray}-0.31 & $<10^{-30}$ & -0.06 & $<10^{-30}$ & 0.15  & $<10^{-30}$ \\
AF/MF, Math/Arts     & \cellcolor{mediumgray}0.59  & $<10^{-30}$ & \cellcolor{lightgray}-0.42 & $<10^{-30}$ & 0.16  & $<10^{-30}$ & \cellcolor{lightgray}-0.25 & $<10^{-30}$ \\
AF/MF, Science/Arts  & \cellcolor{lightgray}-0.20 & $<10^{-30}$ & -0.18 & $<10^{-30}$ & \cellcolor{lightgray}0.22  & $<10^{-30}$ & -0.15 & $<10^{-30}$ \\
MM/MF, P/U$^{\ast}$ - Attitude           & \cellcolor{mediumgray}-0.77 & $<10^{-30}$ & \cellcolor{mediumgray}-0.59 & $<10^{-30}$ & -0.15 & $<10^{-30}$ & \cellcolor{lightgray}-0.44 & $<10^{-30}$ \\
MM/MF, Career/Family & \cellcolor{darkgray}1.11  & $<10^{-30}$ & \cellcolor{lightgray}0.40  & $<10^{-30}$ & \cellcolor{lightgray}0.44  & $<10^{-30}$ & \cellcolor{lightgray}0.42  & $<10^{-30}$ \\
MM/MF, Math/Arts     & \cellcolor{mediumgray}0.62  & $<10^{-30}$ & \cellcolor{mediumgray}0.50  & $<10^{-30}$ & -0.18 & $<10^{-30}$ & \cellcolor{lightgray}-0.49 & $<10^{-30}$ \\
MM/MF, Science/Arts  & 0.18  & $<10^{-30}$ & \cellcolor{lightgray}0.41  & $<10^{-30}$ & -0.19 & $<10^{-30}$ & 0.02  & 0.18        \\ 
\bottomrule
\multicolumn{9}{c}{$^{\ast}$Unpleasant and pleasant attributes used to measure valence and attitudes}\\
\multicolumn{9}{c}{  towards targets  from \citet{greenwald1998measuring}.}

\end{tabular}
\end{small} }
\end{center}
\end{table*}

\begin{table*}
\begin{center}
\captionof{table}{\textbf{CEAT for intersectional groups with sample size $N=1,000$.} We construct 4 types of new CEAT with all pairwise combinations of intersectional groups. We use two intersectional groups as two target social groups. We use 1) pleasant/unpleasant 2) career/family 3) math/arts 4) science/arts as two attribute groups.
Each one of the four experiments with the neural language models is conducted using the same sample of sentences.
We report the CES $d$ and combined $p-value$ $p$. Light, medium, and dark gray shading of combined $d$ values (CES) indicates small, medium, and large effect size respectively. $A\_$ stands for African Americans. $E\_$ stands for European Americans. $M\_$ stands for Mexican Americans. $\_F$ stands for females. $\_M$ stands for males.}

    \resizebox{0.65\textwidth}{!}{%

\begin{small}
\begin{tabular}[h]{lcccccccc}
\label{table:supp-new-same}\\

\textbf{Test} &
  \multicolumn{2}{c}{\textbf{ELMo}} &
  \multicolumn{2}{c}{\textbf{BERT}} &
  \multicolumn{2}{c}{\textbf{GPT}} &
  \multicolumn{2}{c}{\textbf{GPT-2}} \\ \cmidrule(l){2-9} 
                     & $d$   & $p$         & $d$   & $p$         & $d$   & $p$         & $d$   & $p$         \\  \midrule
\hline

                     \hline

EM/EF, P/U$^{\ast}$ - Attitude & \cellcolor{mediumgray}{-0.62}    & $<10^{-30}$   & -0.17    & $<10^{-30}$   & -0.11   & $<10^{-30}$   & \cellcolor{lightgray}{-0.28}    & $<10^{-30}$   \\
      EM/EF, Career/Family                         & \cellcolor{darkgray}{1.07}     & $<10^{-30}$   & \cellcolor{lightgray}{0.40}     & $<10^{-30}$   & \cellcolor{lightgray}{0.27 }   & $<10^{-30}$   & \cellcolor{lightgray}{0.33}     & $<10^{-30}$   \\
       EM/EF, Math/Arts                        & 0.07     & $<10^{-30}$   & 0.12     & $<10^{-30}$   & \cellcolor{lightgray}{0.23}    & $<10^{-30}$   & \cellcolor{lightgray}{-0.22}    & $<10^{-30}$   \\
        EM/EF, Science/Arts                       & \cellcolor{lightgray}{0.21}     & $<10^{-30}$   & -0.04    & $<10^{-30}$   & 0.12    & $<10^{-30}$   & 0.01     &  $<10^{-2}$      \\
            EM/AM, P/U - Attitude                   & \cellcolor{mediumgray}{0.50 }    & $<10^{-30}$   & \cellcolor{lightgray}{0.37}     & $<10^{-30}$   & 0.13    & $<10^{-30}$   & -0.15    & $<10^{-30}$   \\
          EM/AM, Career/Family                     & 0.19     & $<10^{-30}$   & 0.10     & $<10^{-30}$   & \cellcolor{lightgray}{-0.35}   & $<10^{-30}$   & \cellcolor{lightgray}{0.30}     & $<10^{-30}$   \\EM/AM, Math/Arts                              & \cellcolor{lightgray}{-0.47}    & $<10^{-30}$   & -0.18    & $<10^{-30}$   & 0.13    & $<10^{-30}$   & \cellcolor{lightgray}{-0.30 }   & $<10^{-30}$   \\ EM/AM, Science/Arts                               & -0.11    & $<10^{-30}$   & -0.03    &   $<10^{-14}$     & 0.14    & $<10^{-30}$   & \cellcolor{lightgray}{-0.20}    & $<10^{-30}$   \\
           EM/AF, P/U - Attitude                    & \cellcolor{lightgray}{-0.24}    & $<10^{-30}$   & \cellcolor{lightgray}{0.32}     & $<10^{-30}$   & \cellcolor{lightgray}{-0.26}   & $<10^{-30}$   & \cellcolor{lightgray}{-0.26}    & $<10^{-30}$   \\
            EM/AF, Career/Family                   & \cellcolor{darkgray}{1.12}     & $<10^{-30}$   & \cellcolor{lightgray}{0.40}     & $<10^{-30}$   & \cellcolor{lightgray}{0.23}    & $<10^{-30}$   & \cellcolor{mediumgray}{0.66}     & $<10^{-30}$   \\
             EM/AF, Math/Arts                  & 0.07     & $<10^{-30}$   & \cellcolor{lightgray}{0.30}     & $<10^{-30}$   & \cellcolor{lightgray}{0.28}    & $<10^{-30}$   & \cellcolor{mediumgray}{-0.52}    & $<10^{-30}$   \\
              EM/AF, Science/Arts                 & \cellcolor{mediumgray}{0.55}     & $<10^{-30}$   & \cellcolor{lightgray}{0.47 }    & $<10^{-30}$   & \cellcolor{lightgray}{0.21}    & $<10^{-30}$   & \cellcolor{lightgray}{-0.35}    & $<10^{-30}$   \\
               EM/MM, P/U - Attitude                & -0.18    & $<10^{-30}$   & \cellcolor{lightgray}{0.37 }    & $<10^{-30}$   & \cellcolor{lightgray}{-0.36}   & $<10^{-30}$   & -0.12    & $<10^{-30}$   \\
                EM/MM, Career/Family               & -0.08    & $<10^{-30}$   & 0.04     & $<10^{-21}$      & \cellcolor{lightgray}{-0.26 }  & $<10^{-30}$   & \cellcolor{lightgray}{0.23}     & $<10^{-30}$   \\
                 EM/MM, Math/Arts              & -0.08    & $<10^{-30}$   & \cellcolor{lightgray}{-0.22}    & $<10^{-30}$   & \cellcolor{mediumgray}{0.54}    & $<10^{-30}$   & \cellcolor{lightgray}{-0.47 }   & $<10^{-30}$   \\
                   EM/MM, Science/Arts            & 0.16     & $<10^{-30}$   & -0.09    & $<10^{-30}$   & \cellcolor{mediumgray}{0.56}    & $<10^{-30}$   & \cellcolor{lightgray}{-0.45}    & $<10^{-30}$   \\
                   EM/MF, P/U - Attitude            & \cellcolor{mediumgray}{-0.73}    & $<10^{-30}$   & 0.09     & $<10^{-30}$   & \cellcolor{lightgray}{-0.24 }  & $<10^{-30}$   & \cellcolor{lightgray}{-0.24}    & $<10^{-30}$   \\
                    EM/MF, Career/Family           &\cellcolor{darkgray}{ 1.06}     & $<10^{-30}$   & \cellcolor{lightgray}{0.35 }    & $<10^{-30}$   & 0.06    & $<10^{-30}$   & \cellcolor{mediumgray}{0.66 }    & $<10^{-30}$   \\
                    EM/MF,Math/Arts           & \cellcolor{mediumgray}{0.58 }    & $<10^{-30}$   & 0.09     & $<10^{-30}$   & \cellcolor{lightgray}{0.49}    & $<10^{-30}$   & \cellcolor{mediumgray}{-0.61}    & $<10^{-30}$   \\
                     EM/MF, Science/Arts          & \cellcolor{lightgray}{0.24 }    & $<10^{-30}$   & \cellcolor{lightgray}{0.26 }    & $<10^{-30}$   & \cellcolor{lightgray}{0.48}    & $<10^{-30}$   & \cellcolor{lightgray}{-0.43}    & $<10^{-30}$   \\
                      EF/AM, P/U - Attitude         & \cellcolor{darkgray}{0.96}     & $<10^{-30}$   & \cellcolor{mediumgray}{0.51}     & $<10^{-30}$   & \cellcolor{lightgray}{0.24}    & $<10^{-30}$   & 0.12     & $<10^{-30}$   \\
                      EF/AM, Career/Family         & \cellcolor{darkgray}{-1.00}    & $<10^{-30}$   & \cellcolor{lightgray}{-0.31}    & $<10^{-30}$   & \cellcolor{mediumgray}{-0.57}   & $<10^{-30}$   & 0.00     & $0.86$      \\
                      EF/AM, Math/Arts         & \cellcolor{mediumgray}{-0.53}    & $<10^{-30}$   & \cellcolor{lightgray}{-0.30}    & $<10^{-30}$   & -0.10   & $<10^{-30}$   & -0.13    & $<10^{-30}$   \\
                       EF/AM, Science/Arts        & \cellcolor{lightgray}{-0.28}    & $<10^{-30}$   & 0.00     & $0.42$     & 0.03    & $<10^{-13}$      & \cellcolor{lightgray}{-0.24}    & $<10^{-30}$   \\
                        EF/AF, P/U - Attitude       & \cellcolor{lightgray}{0.27}     & $<10^{-30}$   & \cellcolor{lightgray}{0.47}     & $<10^{-30}$   & -0.17   & $<10^{-30}$   & 0.01     & $0.27$      \\
                        EF/AF, Career/Family       & 0.13     & $<10^{-30}$   & 0.01     & $0.037$      & -0.05   & $<10^{-30}$   & \cellcolor{lightgray}{0.45}     & $<10^{-30}$   \\
                        EF/AF, Math/Arts        & 0.00     & $0.85$      & 0.19     & $<10^{-30}$   & 0.06    & $<10^{-30}$   & \cellcolor{lightgray}{-0.40}    & $<10^{-30}$   \\
                        EF/AF, Science/Arts        & \cellcolor{lightgray}{0.34}     & $<10^{-30}$   & \cellcolor{mediumgray}{0.50}     & $<10^{-30}$   & 0.11    & $<10^{-30}$   & \cellcolor{lightgray}{-0.44}    & $<10^{-30}$   \\
                        EF/MM, P/U - Attitude       & \cellcolor{lightgray}{0.47}     & $<10^{-30}$   & \cellcolor{mediumgray}{0.52}     & $<10^{-30}$   & \cellcolor{lightgray}{-0.25}   & $<10^{-30}$   & 0.17     & $<10^{-30}$   \\
                         EF/MM, Career/Family       & \cellcolor{darkgray}{-1.10}    & $<10^{-30}$   & \cellcolor{lightgray}{-0.35 }   & $<10^{-30}$   & \cellcolor{mediumgray}{-0.50}   & $<10^{-30}$   & -0.09    & $<10^{-30}$   \\
                      EF/MM, Math/Arts        & -0.15    & $<10^{-30}$   & \cellcolor{lightgray}{-0.34 }   & $<10^{-30}$   & \cellcolor{lightgray}{0.37}    & $<10^{-30}$   & \cellcolor{lightgray}{-0.26 }   & $<10^{-30}$   \\
                      EF/MM, Science/Arts         & -0.05    & $<10^{-30}$   & -0.06    & $<10^{-30}$   &\cellcolor{lightgray}{ 0.47 }   & $<10^{-30}$   & \cellcolor{mediumgray}{-0.51 }   & $<10^{-30}$   \\
                       EF/MF, P/U - Attitude          & -0.18    & $<10^{-30}$   & \cellcolor{lightgray}{0.26}     & $<10^{-30}$   & -0.14   & $<10^{-30}$   & 0.02     & $<10^{-2}$      \\
                       EF/MF, Career/Family         & -0.13    & $<10^{-30}$   & -0.05    & $<10^{-30}$   & -0.19   & $<10^{-30}$   & \cellcolor{lightgray}{0.46}     & $<10^{-30}$   \\
                     EF/MF, Math/Arts        & \cellcolor{mediumgray}{0.52}     & $<10^{-30}$   & -0.03    & $<10^{-12}$     & \cellcolor{lightgray}{0.32}    & $<10^{-30}$   & \cellcolor{mediumgray}{-0.52}    & $<10^{-30}$   \\
                      EF/MF, Science/Arts          & 0.04     & $<10^{-30}$   & \cellcolor{lightgray}{0.30}     & $<10^{-30}$   & \cellcolor{lightgray}{0.38}    & $<10^{-30}$   & \cellcolor{mediumgray}{-0.52}    & $<10^{-30}$   \\
                        AM/AF, P/U - Attitude      & \cellcolor{mediumgray}{-0.63}    & $<10^{-30}$   & -0.06    & $<10^{-30}$   & \cellcolor{lightgray}{-0.39}   & $<10^{-30}$   & -0.11    & $<10^{-30}$   \\
                         AM/AF, Career/Family       & \cellcolor{darkgray}{1.05}     & $<10^{-30}$   & \cellcolor{lightgray}{0.32}     & $<10^{-30}$   & \cellcolor{mediumgray}{0.53}    & $<10^{-30}$   & \cellcolor{lightgray}{0.41}     & $<10^{-30}$   \\
                      AM/AF, Math/Arts     & \cellcolor{lightgray}{0.49}     & $<10^{-30}$   & \cellcolor{lightgray}{0.48}     & $<10^{-30}$   & 0.16    & $<10^{-30}$   & \cellcolor{lightgray}{-0.24}    & $<10^{-30}$   \\
                    AM/AF, Science/Arts           & \cellcolor{mediumgray}{0.57}     & $<10^{-30}$   & \cellcolor{mediumgray}{0.52}     & $<10^{-30}$   & 0.08    & $<10^{-30}$   & -0.18    & $<10^{-30}$   \\
                    AM/MM, P/U - Attitude           & \cellcolor{mediumgray}{-0.67 }   & $<10^{-30}$   & -0.02    & $<10^{-3}$     & \cellcolor{lightgray}{-0.48}   & $<10^{-30}$   & 0.04     & $<10^{-21}$      \\
                     AM/MM, Career/Family          & \cellcolor{lightgray}{-0.27}    & $<10^{-30}$   & -0.06    & $<10^{-30}$   & 0.13    & $<10^{-30}$   & -0.08    & $<10^{-30}$   \\
                      AM/MM, Math/Arts         & \cellcolor{lightgray}{0.38}     & $<10^{-30}$   & -0.04    & $<10^{-30}$   & \cellcolor{lightgray}{0.45}    & $<10^{-30}$   & -0.15    & $<10^{-30}$   \\
                      AM/MM, Science/Arts         & \cellcolor{lightgray}{0.24}     & $<10^{-30}$   & -0.06    & $<10^{-30}$   & \cellcolor{lightgray}{0.44 }   & $<10^{-30}$   & \cellcolor{lightgray}{-0.27}    & $<10^{-30}$   \\
                     AM/MF, P/U - Attitude          & \cellcolor{darkgray}{-1.03}    & $<10^{-30}$   & \cellcolor{lightgray}{-0.28}    & $<10^{-30}$   & \cellcolor{lightgray}{-0.37}   & $<10^{-30}$   & -0.09    & $<10^{-30}$   \\
                      AM/MF, Career/Family         & \cellcolor{darkgray}{0.98}     & $<10^{-30}$   & \cellcolor{lightgray}{0.25}     & $<10^{-30}$   & \cellcolor{lightgray}{0.38 }   & $<10^{-30}$   & \cellcolor{lightgray}{0.42}     & $<10^{-30}$   \\
                      AM/MF, Math/Arts         & \cellcolor{darkgray}{0.91}     & $<10^{-30}$   & \cellcolor{lightgray}{0.26}     & $<10^{-30}$   & \cellcolor{lightgray}{0.39}    & $<10^{-30}$   & \cellcolor{lightgray}{-0.37}    & $<10^{-30}$   \\
                      AM/MF, Science/Arts         & \cellcolor{lightgray}{0.31}     & $<10^{-30}$   & \cellcolor{lightgray}{0.30}     & $<10^{-30}$   & \cellcolor{lightgray}{0.34 }   & $<10^{-30}$   & \cellcolor{lightgray}{-0.28}    & $<10^{-30}$   \\
                       AF/MM, P/U - Attitude        & 0.09     & $<10^{-30}$   & 0.04     & $<10^{-11}$      & -0.09   & $<10^{-30}$   & 0.16     & $<10^{-30}$   \\
                       AF/MM, Career/Family        & \cellcolor{darkgray}{-1.15}    & $<10^{-30}$   & \cellcolor{lightgray}{-0.36}    & $<10^{-30}$   & \cellcolor{lightgray}{-0.47 }  & $<10^{-30}$   & \cellcolor{lightgray}{-0.47}    & $<10^{-30}$   \\
                      AF/MM, Math/Arts         & -0.14    & $<10^{-30}$   & \cellcolor{mediumgray}{-0.52}    & $<10^{-30}$   & \cellcolor{lightgray}{0.31}    & $<10^{-30}$   & 0.11     & $<10^{-30}$   \\
                      AF/MM, Science/Arts         & \cellcolor{lightgray}{-0.39}    & $<10^{-30}$   & \cellcolor{mediumgray}{-0.56}    & $<10^{-30}$   & \cellcolor{lightgray}{0.35}    & $<10^{-30}$   & -0.10    & $<10^{-30}$   \\
                      AF/MF, P/U - Attitude         & \cellcolor{lightgray}{-0.41}    & $<10^{-30}$   & \cellcolor{lightgray}{-0.23 }   & $<10^{-30}$   & 0.02    & $<10^{-8}$      & 0.02     & $<10^{-3}$      \\
                      AF/MF, Career/Family         & \cellcolor{lightgray}{-0.27}    & $<10^{-30}$   & -0.05    & $<10^{-30}$   & -0.15   & $<10^{-30}$   & 0.06     & $<10^{-30}$   \\
                       AF/MF, Math/Arts        & \cellcolor{lightgray}{0.48}     & $<10^{-30}$   & \cellcolor{lightgray}{-0.22}    & $<10^{-30}$   & \cellcolor{lightgray}{0.26}    & $<10^{-30}$   & -0.17    & $<10^{-30}$   \\
                       AF/MF, Science/Arts        & \cellcolor{lightgray}{-0.29}    & $<10^{-30}$   & \cellcolor{lightgray}{-0.21}    & $<10^{-30}$   & \cellcolor{lightgray}{0.26}    & $<10^{-30}$   & -0.13    & $<10^{-30}$   \\
                      MM/MF, P/U - Attitude         & \cellcolor{mediumgray}{-0.62}    & $<10^{-30}$   & \cellcolor{lightgray}{-0.27}    & $<10^{-30}$   & 0.11    & $<10^{-30}$   & -0.15    & $<10^{-30}$   \\
                    MM/MF, Career/Family           & \cellcolor{darkgray}{1.11}     & $<10^{-30}$   & \cellcolor{lightgray}{0.31}     & $<10^{-30}$   & \cellcolor{lightgray}{0.30}    & $<10^{-30}$   & \cellcolor{lightgray}{0.49}     & $<10^{-30}$   \\
                     MM/MF, Math/Arts          & \cellcolor{mediumgray}{0.63}     & $<10^{-30}$   & \cellcolor{lightgray}{0.30}     & $<10^{-30}$   & -0.04   & $<10^{-30}$   & \cellcolor{lightgray}{-0.25}    & $<10^{-30}$   \\
                    MM/MF, Science/Arts           & 0.09     & $<10^{-30}$   & \cellcolor{lightgray}{0.35 }    & $<10^{-30}$   & -0.08   & $<10^{-30}$   & -0.04    & $<10^{-14}$      \\     
\bottomrule
\multicolumn{9}{c}{$^{\ast}$Unpleasant and pleasant attributes used to measure valence and attitudes}\\
\multicolumn{9}{c}{  towards targets  from \citet{greenwald1998measuring}.}

\end{tabular}
\end{small} }
\end{center}
\end{table*}

\section{Stimuli}
The  stimuli used to represent targets and attributes in CEAT (C1-C10) are taken from Caliskan et al.\cite{caliskan2017semantics}.
We construct four intersection-related CEAT  for African American females and Mexican American females.

When conducting intersection-related  CEAT , 
we use the names from Caliskan et al. \cite{caliskan2017semantics} and Parada et al. \cite{parada2016ethnolinguistic} to represent the target intersectional groups.  Caliskan et al.'s WEAT provides the female and male names of African Americans and European Americans from the first Implicit Association Test in 1998 \cite{greenwald1998measuring}. Parada et al. provide the female and male names of Mexican Americans \cite{parada2016ethnolinguistic}. To determine and verify the gender of names, we use three gender checkers \cite{huang2019gender}. We only use the name as a target word in our experiments, if the name is categorized to belong to the same gender by all of the three checkers. Human subjects provide the validation set of intersectional attributes with ground truth information \cite{ghavami2013intersectional}. We use this validation set for evaluating the intersection-related CEAT, IBD and EIBD experiments.
To follow the order of stereotype-congruity, we use European American males as the second target group and use the attributes associated with their intersectional biases as the second attribute set in intersection-related CEAT. There are only three emergent intersectional biases associated with European American males in the validation set, which doesn't provide a sufficient number of stimuli. A small set of stimuli does not satisfy the requirements for generating statistically significant concept representation and WEATs. Related stimuli details are discussed in the dataset and stimuli sections of the main paper. In addition, if the size of the first attribute set is smaller than that of the attributes of European American males, we randomly select an equal number of attributes associated with the intersectional biases of European American males. WEAT requires equal-sized sets of attributes.


\subsection{CEAT I1}
We use the frequent given names of African American females and European American males as two target social groups and use the attributes associated with the intersectional biases of African American females and attributes associated with the intersectional biases of European American males as the two attribute groups.

Since `assertive' is associated with both African American females and European American males, we do not include it in this test.

\begin{itemize}
    \item \textbf{African American females}:  Aisha, Keisha, Lakisha, Latisha, Latoya, Malika, Nichelle, Shereen, Tamika, Tanisha, Yolanda, Yvette
    \item \textbf{European American males}: Andrew, Brad, Frank, Geoffrey, Jack, Jonathan, Josh, Matthew, Neil, Peter, Roger, Stephen
     \item \textbf{Intersectional biases of African American females}: aggressive, athletic, bigbutt, confident, darkskinned, fried-chicken, ghetto, loud, overweight, promiscuous, unfeminine, unintelligent, unrefined
    \item \textbf{Intersectional biases of European American males}: all-American, arrogant, attractive, blond, high-status, intelligent, leader, privileged, racist, rich, sexist, successful, tall
\end{itemize}

\subsection{CEAT I2}
We use the frequent given names of African American females and European American males as  two target groups. We use attributes associated with emergent intersectional biases of African American females and attributes associated with  intersectional biases of European American males as two attribute groups.

Since `assertive' is associated with emergent intersectional bias of African American females and intersectional bias of European American males, we do not include it in this test.

\begin{itemize}
    \item \textbf{African American females}: Aisha, Keisha, Lakisha, Latisha, Latoya, Malika, Nichelle, Shereen, Tamika, Tanisha, Yolanda, Yvette
    \item \textbf{European American males}: Andrew, Brad, Frank, Geoffrey, Jack, Jonathan, Josh, Matthew, Neil, Peter, Roger, Stephen
    \item \textbf{Emergent intersectional biases of African American females}: aggressive, bigbutt, confident, darkskinned, fried-chicken, overweight, promiscuous, unfeminine
    \item \textbf{Intersectional biases of European American males}: arrogant, blond, high-status, intelligent, racist, rich, successful, tall
\end{itemize}

\subsection{CEAT I3}
We use the frequent given names of Mexican American females and European American males as the target groups and the words associated with their intersectional biases as the attribute groups.

Since `attractive' is associated with intersectional biases of both Mexican American females and European American males, we do not include it in this test.

\begin{itemize}
    \item \textbf{Mexican American females}:  Adriana, Alejandra, Alma, Brenda, Carolina, Iliana, Karina, Liset, Maria, Mayra, Sonia, Yesenia
    \item \textbf{European American males}: Andrew, Brad, Frank, Geoffrey, Jack, Jonathan, Josh, Matthew, Neil, Peter, Roger, Stephen
    \item \textbf{Intersectional biases of Mexican American females}: cook, curvy, darkskinned, feisty, hardworker, loud, maids, promiscuous, sexy, short, uneducated, unintelligent
    \item \textbf{Intersectional biases of European American males}: all-American, arrogant, blond, high-status, intelligent, leader, privileged, racist, rich, sexist, successful, tall
\end{itemize}

\subsection{CEAT I4}
We use the frequent given names of Mexican American females and European American males as target groups. We use words associated with the emergent intersectional biases of Mexican American females and words associated with the intersectional biases of European American males as the two attribute groups.

\begin{itemize}
    \item \textbf{Mexican American females}:  Adriana, Alejandra, Alma, Brenda, Carolina, Iliana, Karina, Liset, Maria, Mayra, Sonia, Yesenia
    \item \textbf{European American males}: Andrew, Brad, Frank, Geoffrey, Jack, Jonathan, Josh, Matthew, Neil, Peter, Roger, Stephen
    \item \textbf{Emergent intersectional biases of Mexican American females}: cook, curvy, feisty, maids, promiscuous, sexy
    \item \textbf{Intersectional biases of European American males}: arrogant, assertive, intelligent, rich, successful, tall
\end{itemize}

\subsection{IBD and EIBD}
We detect the attributes associated with the intersectional biases and emergent intersectional biases of African American females and Mexican American females in GloVe  SWE. We assume that there are three subcategories under the race category (African American, Mexican American, European American) and two subcategories under the gender category (female, male). We use the frequent given names to represent each intersectional group. Again, we note that, in future work we'd generalize this work to $n$ subcategories under each category. Further, in future work, instead of categorizing people into social groups, we'd like to explore representing individuals in social data with continuous real-valued variables as opposed to associating them with category labels.

\begin{itemize}
        \item \textbf{African American females}:  Aisha, Keisha, Lakisha, Latisha, Latoya, Malika, Nichelle, Shereen, Tamika, Tanisha, Yolanda, Yvette
        \item \textbf{African American males}: Alonzo, Alphonse, Hakim, Jamal, Jamel, Jerome, Leroy, Lionel, Marcellus, Terrence, Tyrone, Wardell
                \item \textbf{European American females}: Carrie, Colleen, Ellen, Emily, Heather, Katie, Megan, Melanie, Nancy, Rachel, Sarah,\\Stephanie
        \item \textbf{European American males}: Andrew, Brad, Frank, Geoffrey, Jack, Jonathan, Josh, Matthew, Neil, Peter, Roger, Stephen
        \item \textbf{Mexican American females}:  Adriana, Alejandra, Alma, Brenda, Carolina, Iliana, Karina, Liset, Maria, Mayra, Sonia, Yesenia
        \item \textbf{Mexican American males}: Alberto, Alejandro, Alfredo, Antonio, César, Jesús, José, Juan, Miguel, Pedro, Rigoberto, Rogelio

\end{itemize}

We list all the attributes used in validation experiments. These are stimuli associated with different social groups and a set of random words that are not expected to be associated with social groups. These random attributes are borrowed from the insect target category of the `universally accepted stereotypes' IAT, which is a baseline WEAT. We use 98 words in total in the validation experiments.

We also list the probability of random chance of correct classification in parentheses next to each social group. The goal here is to present the success rate of the machine learning detection method in a meaningful way. To calculate the probability of random chance of correct classification, we treat the detection task as if this were a one-vs-all verification task, which is one type of a classification task, that aims to verify the membership of a data point in the main class. All the other data points from the open world, that don't belong to the main class we are testing for, can belong to any other class. However, in reality, our detection method is a semi-supervised non-parametric prediction method which relies on the threshold $t$.
Some of the attributes of some social groups overlap. For example, `assertive' is associated with emergent intersectional biases of African American females and also associated with emergent intersectional biases of European American males. As a result, the sum of all random chances of different groups is not 100\%.
\begin{itemize}
    \item \textbf{Biases of Females}: attractive, caring, dependent, emotional, feminine, jealous, manipulative, materialistic, motherly, petite, soft, submissive, talkative (random chance of correct classification: 13.3\%)
    \item \textbf{Biases of Males}: aggressive, ambitious, arrogant, fixer-upper, high-status, intelligent, leader, messy, provider, respected, sexist, tall, unfaithful (random chance of correct classification: 13.3\%)
    
    \item \textbf{Biases of African Americans}: athletic, criminals, dangerous, gangsters, ghetto, lazy, loud, poor, tall, uneducated, unrefined, violent (random chance of correct classification: 12.2\%)
    \item \textbf{Emergent Intersectional Biases of African American Females}: aggressive, assertive, bigbutt, confident, \\darkskinned, fried-chicken, overweight, promiscuous, unfeminine (random chance of correct classification: 9.2\%)
    \item \textbf{Intersectional Biases of African American Females}: aggressive, assertive, athletic, bigbutt, confident, darkskinned, fried-chicken, ghetto, loud, overweight, promiscuous, unfeminine, unintelligent, unrefined (random chance of correct classification: 14.3\%)
    \item \textbf{Emergent Intersectional Biases of African American Males}: darkskinned, hypersexual, rapper (random chance of correct classification: 3.1\%)
    \item \textbf{Intersectional Biases of African American Males}: athletic, criminals, dangerous, darkskinned, gangsters, hypersexual, lazy, loud, poor, rapper, tall, unintelligent, violent (random chance of correct classification: 13.3\%)

    \item \textbf{Biases of European Americans}: all-American, arrogant, attractive, blond, blue-eyes, high-status, ignorant, intelligent, overweight, patronizing, privileged, racist, red-neck, rich, tall (random chance of correct classification: 15.3\%)
      \item \textbf{Emergent Intersectional Biases of European American Females}: ditsy (random chance of correct classification: 1.0\%)
    \item \textbf{Intersectional Biases of European American Females}: arrogant, attractive, blond, ditsy, emotional, feminine, high-status, intelligent, materialistic, petite, racist, rich, submissive, tall (random chance of correct classification: 14.3\%)
    \item \textbf{Emergent Intersectional Biases of European American Males}: assertive, educated, successful (random chance of correct classification: 3.1\%)
    \item \textbf{Intersectional Biases of European American Males}: all-American, arrogant, assertive, attractive, blond, educated, high-status, intelligent, leader, privileged, racist, rich, sexist, successful, tall (random chance of correct classification: 15.3\%)

   \item \textbf{Biases of Mexican Americans}: darkskinned, day-laborer, family-oriented, gangster, hardworker, illegal-immigrant, lazy, loud, macho, overweight, poor, short, uneducated, unintelligent (random chance of correct classification: 14.3\%)
    \item \textbf{Emergent Intersectional Biases of Mexican American Females}: cook, curvy, feisty, maids, promiscuous, sexy (random chance of correct classification: 6.1\%)
    \item \textbf{Intersectional Biases of Mexican American Females}: attractive, cook, curvy, darkskinned, feisty, hardworker, loud, maids, promiscuous, sexy, short, uneducated, unintelligent (random chance of correct classification: 13.3\%)
    \item \textbf{Emergent Intersectional Biases of Mexican American Males}: drunks, jealous, promiscuous, violent (random chance of correct classification: 4.1\%)
    \item \textbf{Intersectional Biases of Mexican American Males}: aggressive, arrogant, darkskinned, day-laborer, drunks, hardworker, illegal-immigrant, jealous, macho, poor, promiscuous, short, uneducated, unintelligent, violent (random chance of correct classification: 15.3\%)
 
    \item   \textbf{Random (Insects)}: ant, bedbug, bee, beetle, blackfly, caterpillar, centipede, cockroach, cricket, dragonfly, flea, fly, gnat, hornet, horsefly, locust, maggot, mosquito, moth, roach, spider, tarantula, termite, wasp, weevil (random chance of correct classification: 25.5\%)
\end{itemize}

\section{Open Source Code, Data, and Documentation}
\url{https://github.com/weiguowilliam/CEAT} is the link to our open source git repository. Code and links to datasets are available in the project repository. In addition, answers to frequently asked questions about the details of extracting the contextualized word embeddings are documented. The extracted embeddings for the stimuli take up approximately $\sim50GB$ memory.


\newpage
\bibliography{ceat}
\end{document}